\documentclass[]{revtex4-2}

\usepackage[utf8]{inputenc} 
\usepackage[T1]{fontenc}  

\usepackage{verbatim} 

\usepackage{graphicx}
\usepackage{dcolumn}

\usepackage{amstext,amsmath,amssymb,amsfonts}
\usepackage{bm}
\usepackage{stmaryrd}

\usepackage{lmodern}
\usepackage{mathtools}
\DeclareMathAlphabet{\mathpzc}{OT1}{pzc}{m}{it} 
\DeclareFontFamily{OT1}{pzc}{} 
\DeclareFontShape{OT1}{pzc}{m}{it}{<-> s * [1.100] pzcmi7t}{} 
\DeclareMathAlphabet{\mathpzc}{OT1}{pzc}{m}{it} 

\usepackage{float}
\usepackage{hyperref}

\usepackage{color}
\definecolor{lightblue}{rgb}{0.2,0.2,0.7}
\definecolor{darkblue}{rgb}{0,0.25,0.5}
\definecolor{redbrown}{rgb}{0.875,0.25,0.125}
\definecolor{darkgreen}{rgb}{0,0.5,0}

\newcommand{\braket}[2]{\ensuremath{\langle  #1, #2  \rangle}}
\renewcommand{\b}[1]{\ensuremath{\mathbf{#1}}}

\newcommand{\ee}{\ensuremath{\text{ee}}}
\newcommand{\tne}{\ensuremath{\text{ne}}}

\renewcommand{\d}{\ensuremath{\text{d}}}

\renewcommand{\L}{\ensuremath{\text{L}}}
\newcommand{\R}{\ensuremath{\text{R}}}
\newcommand{\calh}{\ensuremath{\mathpzc{h}}}
\newcommand{\calw}{\ensuremath{\mathpzc{w}}}
\newcommand*{\llbrace}{\{\!\{}
\newcommand*{\rrbrace}{\}\!\}}

\makeatletter
\renewcommand\paragraph{\@startsection{paragraph}{4}{\z@}%
  {-3.25ex\@plus -1ex \@minus -.2ex}%
  {1.5ex \@plus .2ex}%
  {\normalfont\normalsize\bfseries}}
\makeatother

\begin{document}

\title{Strictly localized orbitals from spatial partitioning with the discontinuous Galerkin method}

\author{César Feniou}
\affiliation{Qubit Pharmaceuticals, Advanced Research Department, F-75014 Paris, France}
\author{Julien Toulouse}
\email{julien.toulouse@sorbonne-universite.fr}
\affiliation{Laboratoire de Chimie Th\'eorique, Sorbonne Universit\'e and CNRS, F-75005 Paris, France}

\date{August 13, 2026}

\begin{abstract}
We present a rigorous electronic-structure theory of strictly localized orbitals associated with a spatial partition of the one-electron Hilbert space that remains well defined in the complete basis-set limit. Each strictly localized orbital is supported on a spatial domain and may be discontinuous at domain interfaces. Using the interior-penalty discontinuous Galerkin method, these strictly localized orbitals can be employed in variational electronic-structure calculations despite their discontinuities at domain interfaces. As a proof of concept, we present numerical illustrations on one-dimensional diatomic model systems. They show that variational calculations can be carried out in a basis of strictly localized orbitals while maintaining good agreement with conventional calculations. Moreover, these strictly localized orbitals naturally lead to chemically intuitive representations of many-electron wave functions in the spirit of valence-bond theory.
\end{abstract}

\maketitle

\section{Introduction}

Orbitals play a central role in quantum chemistry, providing a practical representation of many-electron wave functions and enabling chemically intuitive descriptions of electronic structures. However, orbitals are not uniquely defined. In fact, any invertible transformation of the orbitals leaves a full-configuration-interaction wave function invariant, leading to a wide range of possible definitions of orbitals with distinct computational and interpretational properties. Among these, localized orbitals (see, e.g., Ref.~\onlinecite{BenEvaLeiAnd-INC-21}) are particularly valuable, as they offer a compact and chemically transparent picture of bonding while also improving computational efficiency through sparsity and reduced entanglement.

A variety of schemes to define localized orbitals have been developed over the years (see, e.g., Ref.~\onlinecite{HoyJor-CR-16}). A first family of schemes consists in variational optimization of localization functionals. In this family, the three most used schemes are the one of Boys which minimizes the sum of the orbital spreads~\cite{Boy-RMP-60}, the one of Edmiston and Ruedenberg which maximizes the sum of the orbital self-repulsion energies~\cite{EdmRue-RMP-63}, and the one of Pipek and Mezey which maximizes the sum of orbital partial charges on nuclei~\cite{PipMez-JCP-89}. For periodic solids, the equivalent of the Boys localized orbitals are called maximally localized Wannier functions~\cite{MarMosYatSouVan-RMP-12}. A second family of schemes consists in starting from a set of localized orbitals that do not minimize the energy and find a small transformation of these orbitals that minimizes the energy~\cite{Dau-CPL-74,RubPovMalRei-JCP-97,LiLiSuoLiu-ACR-14}. A third family of schemes consists in defining nonorthogonal localized orbitals by expanding them on localized basis functions, such as Gaussian-type orbital (GTO) basis functions, centered on a single atom or a limited set of atoms of the molecule~\cite{StoWagPre-TCA-80,KhaHeaBel-JCP-06,SirGenCivPieGhi-TCA-07} (see also Ref.~\onlinecite{MccFauZhuOgoQiuWhiRabLin-NJP-20,FauWuLin-RMS-22}). This last family of nonorthogonal localized orbitals are used in particular in classical valence-bond (VB) theory~\cite{ShaHib-BOOK-08} but they have the serious drawback that their definition is tied to the basis set and tend to lose their meaning for large basis sets.

None of the previously mentioned schemes give orbitals strictly localized on a single atom or a single fragment of the molecule, where by strictly localized orbitals we mean orbitals which are strictly non-zero only on the spatial domain defining a single atom or a single fragment of the molecule, a property that may be desirable for both computational and interpretational purposes. Mayer~\cite{May-CJP-96,May-IJQC-14} proposed to define such strictly localized orbitals based on a spatial partitioning such as the one provided by the quantum theory of atoms in molecules (QTAIM)~\cite{Bad-BOOK-94}. The idea is to generate localized orbitals for an atom by truncating canonical molecular orbitals to the spatial domain corresponding to the QTAIM basin of this atom, and then orthonormalize them. These so-called effective atomic orbitals have been calculated for a few molecules and have the nice feature of reconciling Hilbert-space analysis with real-space analysis~\cite{RamSalMay-JCP-13}. In this last study, the effective atomic orbitals were calculated at posteriori from canonical molecular orbitals. Since these strictly localized orbitals are discontinuous on the boundaries of atomic basins, they have infinite kinetic energy and thus do not constitute a proper orbital basis to straightforwardly calculate the energy, which may be one reason for which they have not been adopted yet. It has been proposed alternative ways of defining orbitals either based on QTAIM~\cite{CioLia-JCP-98,Pon-JMC-98,TiaFraBlaMacSirMar-PCCP-11} or on a fuzzy definition of atoms in a molecule~\cite{MaySal-JCP-09,May-IJQC-14}, but the resulting orbitals are no longer strictly localized.

In applied mathematics, discontinuous Galerkin methods are a class of numerical methods to solve differential equations using basis functions that are only piecewise continuous, such as discontinuous piecewise polynomials (see, e.g., Ref.~\onlinecite{DipErn-BOOK-12}). The discontinuous Galerkin approach was first introduced in electronic-structure theory by Lin and coworkers~\cite{LinLuYinE-JCP-12,HuLinYan-JCP-15,ZhaLinHuYanPas-JCP-17,LiLin-SIAM-19}. These authors used the interior-penalty discontinuous Galerkin method to generate a reduced basis set made of adaptive strictly localized basis functions in order to speed up periodic Kohn-Sham (KS) calculations.

In the present work, we combine the idea of strictly localized orbitals based on a chemically relevant spatial partitioning and the interior-penalty discontinuous Galerkin method. By allowing one to perform energy minimization with discontinuous strictly localized basis functions, the interior-penalty discontinuous Galerkin method provides a rigorous variational formulation for the effective atomic orbitals of Mayer and coworkers~\cite{May-CJP-96,RamSalMay-JCP-13,May-IJQC-14}. More generally, the interior-penalty discontinuous Galerkin method allows one to formulate a well-defined electronic-structure theory of strictly localized orbitals in quantum chemistry in the infinite-dimensional setting (i.e., in the limit of a complete basis set), which is thus properly independent from the choice of basis set. 

The paper is organized as follows. In Section~\ref{sec:theory}, we provide the theory of strictly localized orbitals based on the interior-penalty discontinuous Galerkin method. In Section~\ref{sec:numerical}, we give numerical illustrations on one-dimensional (1D) model systems. Finally, Section~\ref{sec:conclusion} contains our conclusions.
Atomic units are used throughout this work.

\section{Theory}
\label{sec:theory}

\subsection{Exact molecular ground-state electronic-structure theory}

We start by a brief review of the mathematical formulation of the exact molecular electronic-structure theory (for a general mathematical review, see e.g. Ref.~\onlinecite{CanDefKutLebMad-BOOK-03}). We consider $N$ electrons in a molecule with position-spin coordinates $\b{x}_i=(\b{r}_i,\sigma_i)\in \mathbb{R}^3 \times \Sigma$ where $\Sigma = \{ \uparrow,\downarrow \}$ is the set of spin coordinates. The one-electron Hilbert space is ${\calh} = L^2(\mathbb{R}^3 \times\Sigma ,\mathbb{C})$, i.e. the space of square-integrable functions, with inner product denoted as $\braket{\cdot}{\cdot}$. The spin-orbital space ${\calw} \subset \calh$ is taken as the first-order Sobolev space
\begin{equation}
\textstyle
{\calw} = H^1 (\mathbb{R}^3 \times\Sigma,\mathbb{C}) = \{ \phi \in \calh \;|\; \nabla \phi \in \calh^3 \},
\label{w}
\end{equation}
representing the space of square-integrable functions with gradient also square integrable.
This spin-orbital space allows one to define the action of the one-electron kinetic-energy operator $\hat{t}=-(1/2)\Delta$ via its sesquilinear form for any two spin-orbitals $\phi,\psi\in {\calw}$ as
\begin{equation}
\braket{\phi}{\hat{t}\; \psi} = \frac{1}{2} \braket{\nabla\phi}{\nabla\psi} = \frac{1}{2} \int_{\mathbb{R}^3\times\Sigma} \nabla \phi^*(\b{x}) \cdot \nabla \psi(\b{x}) \d \b{x}.
\label{phitpsi}
\end{equation}
In this sense, ${\calw}$ is the largest spin-orbital space ensuring finite kinetic energy (the so-called form domain of the operator $t$). In the special case where $\psi$ is in the second-order Sobolev space (the domain of the operator $t$), i.e. $\psi \in H^2 (\mathbb{R}^3 \times\Sigma,\mathbb{C}) = \{ \phi \in \calh \;|\; \Delta \phi \in \calh \} \subset {\calw}$, we recover by integration by parts the familiar Laplacian form of the kinetic-energy operator: for any $\phi\in {\calw}$,  $\braket{\phi}{\hat{t}\; \psi} = -(1/2) \braket{\phi}{\Delta\psi}$.

The $N$-electron Hilbert space is given by the $N$-fold antisymmetrized tensor product of one-electron Hilbert spaces, i.e. ${\cal H} = \bigwedge^N {\calh}$ with inner product still denoted as $\braket{\cdot}{\cdot}$. The $N$-electron molecular Born-Oppenheimer non-relativistic Hamiltonian is
\begin{eqnarray}
\hat{H} = \hat{T} + \hat{V}_{\tne} + \hat{W}_\ee,
\label{H}
\end{eqnarray}
where $\hat{T}=\sum_{i=1}^{N} \hat{t}_i$ is the total kinetic-energy operator, $\hat{V}_{\tne}=\sum_{i=1}^{N} v_{\tne}(\b{r}_i)$ with $v_{\tne}(\b{r}_i)=- \sum_{k=1}^{M_\text{n}} Z_{k}/|\b{r}_i - \b{R}_{k}|$ is the nuclei-electron interaction depending of the positions $\{\b{R}_{k}\}$ and charges $\{Z_{k}\}$ of the $M_\text{n}$ nuclei, and $\hat{W}_\ee=\sum_{1\leq i<j\leq N} 1/|\b{r}_i -\b{r}_j|$ is the electron-electron interaction. According to the variational theorem, the ground-state energy $E_0$ can be expressed as an infimum
\begin{equation}
E_{0} = \inf_{\Psi \in {\cal W}} \braket{\Psi}{\hat{H} \Psi},
\label{E0N}
\end{equation}
where the search is over the space of admissible normalized $N$-electron wave functions $\Psi$ 
\begin{equation}
\textstyle
{\cal W} = \Big\{ \Psi \in \bigwedge^N {\calw} \, \;|\; \braket{\Psi}{\Psi}=1 \Big\} \subset {\cal H},
\label{W}
\end{equation}
which is the largest space of normalized $N$-electron wave functions having finite kinetic energy and in fact finite expectation value of the whole Hamiltonian $\hat{H}$.

Let us consider now a (complete) spin-orbital orthonormal basis $\{\psi_i \}_{i \in \mathbb{N}} \subset \calw$ of the one-electron Hilbert space $\calh$.
The space of admissible normalized $N$-electron wave functions can then be explicitly written  as
\begin{eqnarray}
{\cal W} &=& \Big\{ \Psi = \sum_{I \in {\cal D}} C_{I} \; \psi_{I_1} \wedge \psi_{I_2} \wedge \cdots \wedge \psi_{I_N}  \;|\; \braket{\Psi}{\Psi}=1 ;
(C_{I} )_{I \in {\cal D}} \in \mathbb{C}^{\cal D} \Big\},
\end{eqnarray}
where $\psi_{I_1} \wedge \psi_{I_2} \wedge \cdots \wedge \psi_{I_N}$ is a Slater determinant (i.e., a normalized antisymmetrized product of $N$ spin-orbitals), ${\cal D}$ is the set of ordered spin-orbital labels of all $N$-electron Slater determinants 
\begin{equation}
{\cal D} = \left\{ I=(I_1,I_2,...,I_N) \in \mathbb{N}^N \;|\; I_1<I_2<\cdots<I_N \right\},
\label{DN}
\end{equation}
and $(C_{I})_{I \in {\cal D}}$ is a vector of complex coefficients.
The choice of the spin-orbital basis $\{\psi_i \}_{i \in \mathbb{N}}$ is completely arbitrary: any wave function $\Psi \in {\cal W}$ can be expressed with any spin-orbital basis (but of course with different coefficients $C_I$). One can exploit this freedom for defining a basis of localized spin-orbitals. As we will see in the next section, part of this arbitrariness in the definition of the spin-orbitals is lost when restrictions are introduced on the variational wave-function space ${\cal W}$.

\subsection{Constraints on orbitals by restrictions on the variational wave-function space}
\label{sec:restrict_wf_space}

In variational electronic-structure methods, we usually consider a subset ${\cal D}' \subset {\cal D}$ of $N$-electron Slater determinants and a possibly constrained set of coefficients ${\cal C}' \subset \mathbb{C}^{\cal D'}$, and the associated variational energy is obtained as
\begin{equation}
E_0'= \inf_{\Psi \in {\cal W}'} E[\Psi],
\label{ED}
\end{equation}
where $E[\Psi]$ is a functional of $\Psi$ and ${\cal W}' \subset {\cal W}$ is the subspace of admissible normalized $N$-electron wave functions
\begin{eqnarray}
{\cal W}' &=& \Big\{ \Psi = \sum_{I \in {\cal D}'} C_{I} \; \phi_{I_1} \wedge \phi_{I_2} \wedge \cdots \wedge \phi_{I_N}  \;|\; \braket{\Psi}{\Psi}=1;
(C_I)_{I \in {\cal D}'} \in {\cal C}';
\{\phi_{i} \}_{i \in \mathbb{N}} = \hat{U} \{ \psi_{i} \}_{i \in \mathbb{N}}, \; \hat{U}\in {\cal U}({\calh})
\Big\},
\nonumber\\
\label{WD}
\end{eqnarray}
where ${\cal U}({\calh})$ designates the set of all unitary transformations on ${\calh}$. Thus, in the minimization in Eq.~(\ref{ED}), the coefficients $(C_I)_{I \in {\cal D}'}$ of the Slater determinants are varied, as well as the spin-orbitals $\{\phi_i \}_{i \in \mathbb{N}}$ of these Slater determinants via unitary transformations $\hat{U}$ of the starting spin-orbital orthonormal basis $\{\psi_i \}_{i \in \mathbb{N}}$. This encompasses variational wave-function methods for which $E[\Psi] = \braket{\Psi}{\hat{H} \Psi}$, such as Hartree-Fock (HF) or multi-configuration self-consistent-field (MCSCF) in a complete basis set~\cite{Lew-ARMA-04}, but also density-functional-theory (DFT) approaches (see, e.g., Ref.~\onlinecite{Tou-INC-23}) for which $E[\Psi]=\braket{\Psi}{\hat{H}' \Psi}+{\cal E}[\rho_\Psi]$ where $\hat{H}'$ is an effective Hamiltonian and ${\cal E}[\rho_\Psi]$ is a functional of the one-electron density $\rho_\Psi$ of $\Psi$. For DFT approaches, $E_0'$ can in principle be equal to the exact ground-state energy $E_0$. The orthonormality condition on the set of spin-orbitals $\{ \phi_i \}_{i\in \mathbb{N}}$ can also be relaxed (see, e.g., Ref.~\onlinecite{Ols-JCP-15}) by just requiring the transformation $\hat{U}$ in Eq.~(\ref{WD}) to be invertible and not unitary. The obtained spin-orbitals $\{ \phi_i \}_{i \in \mathbb{N}}$ then form a nonorthogonal basis, also called a Riesz basis in the infinite-dimensional setting~\cite{Chr-BOOK-03,Hei-BOOK-11}.

For a given choice of determinant set ${\cal D}'$, a minimizing wave function $\Psi_0'$ in Eq.~(\ref{ED}), assuming it exists, is written with two subsets of occupied spin-orbitals: 
\begin{itemize}
\item[(1)] a set of $M_{\text{c}}$ \textit{core} (or inactive) spin-orbitals ($M_\text{c} \leq N$) which are occupied in all Slater determinants listed in ${\cal D}'$
\begin{equation}
{\cal O}_\text{c} = \{ \phi_i \;|\; \forall I \in {\cal D}', i \in I  \},
\label{}
\end{equation}
where $i \in I$ means that the label $i$ is one of the spin-orbital labels of $I$; 

\item[(2)] a set of $M_{\text{a}}$ \textit{active} spin-orbitals which are occupied in at least one Slater determinant listed in ${\cal D}'$ and unoccupied in at least another one
\begin{equation}
{\cal O}_\text{a} = \{ \phi_i \;|\; \exists\; I,J \in {\cal D}',  i  \in I, i \not\in J \}.
\label{}
\end{equation}
\end{itemize}
It is always possible to choose active spin-orbitals that are orthogonal to the core spin-orbitals. In this case, we have the following orthogonal direct-sum decomposition of the one-electron space ${\calw}$
\begin{equation}
{\calw} = {\calw}_{\text{c}} \oplus {\calw}_{\text{a}} \oplus {\calw}_{\text{v}},
\label{wdecompCAV}
\end{equation}
where ${\calw}_{\text{c}} = \text{span}({\cal O}_\text{c})$ is the core spin-orbital space, ${\calw}_{\text{a}} = \text{span}({\cal O}_\text{a})$ is the active spin-orbital space, and ${\calw}_{\text{v}}$ is the \textit{virtual} (or secondary) spin-orbital space (the orthogonal complement of ${\calw}_{\text{c}} \oplus {\calw}_{\text{a}}$ in ${\calw}$).

There is full freedom in choosing a spin-orbital basis within the core space ${\calw}_{\text{c}}$ and a spin-orbital basis within the virtual space ${\calw}_{\text{v}}$. The situation within the active space ${\calw}_{\text{a}}$ is more involved and depends on the choice of the determinant set ${\cal D}'$ and coefficient set ${\cal C}'$. For general sets ${\cal D}'$ and ${\cal C}'$, the energy minimization in Eq.~(\ref{ED}) imposes some constraints on the active spin-orbitals.
There are two important special cases in which the situation for the active space is simple: 
\begin{itemize}
\item Single-determinant methods (such as HF or KS): one considers only one Slater determinant, i.e. ${\cal D}' = \{(1,2,...,N)\}$. There is no active spin-orbitals, i.e. ${\cal O_\text{a}}=\emptyset$ and ${\calw}_{\text{a}}= \{0\}$.
\item Complete-active-space self-consistent field (CASSCF): one considers linear combinations of all the Slater determinants generated by occupying the active spin-orbitals with $N-M_{\text{c}}$ electrons while keeping the $M_{\text{c}}$ core spin-orbitals always occupied, i.e. 
${\cal D}' =\{ I \in {\cal D} \;|\; I_1=1, I_2=2, ..., I_{M_{\text{c}}} = M_{\text{c}}, \; \text{and} \; I_N \leq {M_{\text{c}}+M_{\text{a}}}  \}$ and ${\cal C}' = \mathbb{C}^{{\cal D}'}$. Then, the energy minimization in Eq.~(\ref{ED}) does not impose any constraints on the active spin-orbitals, i.e. we have full freedom in choosing a spin-orbital basis within the active space ${\calw}_{\text{a}}$.
\end{itemize}

We thus see that imposing restrictions on the variational wave-function space reduces the arbitrariness in the definition of the spin-orbitals via the decomposition in Eq.~\eqref{wdecompCAV}. If one does not want to raise the energy, only the remaining freedom could be exploited to define localized spin-orbitals. Sometimes, the variational wave-function ansatz tends to automatically give fairly localized active spin-orbitals after energy minimization, such as in the generalized VB~\cite{BobGod-INC-77,GodDunHunJef-ACR-73} and spin-coupled VB~\cite{CooGerRai-CR-91} theories. Note of course that in the limit of the exact theory, i.e. ${\cal D}'={\cal D}$ and ${\cal C}' = \mathbb{C}^{{\cal D}}$, all the spin-orbitals are active, i.e. ${\calw} = {\calw}_{\text{a}}$, and we recover total freedom in choosing a spin-orbital basis.

\subsection{Localized orbitals from spatial partitioning and the discontinuous Galerkin method}

Inspired by the work of Mayer and coworkers~\cite{May-CJP-96,RamSalMay-JCP-13,May-IJQC-14} and the work of Lin and coworkers~\cite{LinLuYinE-JCP-12,HuLinYan-JCP-15,ZhaLinHuYanPas-JCP-17,LiLin-SIAM-19}, we propose to rigorously define subspaces of localized spin-orbitals based on a spatial partitioning and the interior-penalty discontinuous Galerkin method. We thus consider a partition ${\cal P}$ of $\mathbb{R}^3$
\begin{equation}
{\cal P} = \left\{\Omega_1, \Omega_2, ...,\Omega_{M_{\cal{P}}} \right\},
\label{}
\end{equation}
such that $\{\Omega_p\}_{p=1,...,M_{\cal{P}}}$ are $M_{\cal{P}}$ open disjoint domains of $\mathbb{R}^3$ with $\bigcup_{p=1}^{M_{\cal{P}}} \overline{\Omega}_p = \mathbb{R}^3$ and $\overline{\Omega}_p$ designates the topological closure of $\Omega_p$. The set of distinct interfaces between two domains is 
\begin{equation}
{\cal S}_{\cal P}=\left\{ S= \partial \Omega_1 \cap \partial \Omega_2 \; | \; \Omega_1, \Omega_2 \in {\cal P},\; \Omega_1 \not=\Omega_2 \right\}, 
\label{}
\end{equation}
where $\partial \Omega$ designates the boundary of the domain $\Omega$. For example, for a given molecule, it could be the atomic partition realized by the Voronoi polyhedra~\cite{Med-JCP-86} generated by the nuclear positions (appropriately modified to account for the different sizes of atoms~\cite{Bec-JCP-88b,RouPeeAls-JMS-01}), or the QTAIM atomic partition~\cite{Bad-BOOK-94}, where the space is divided into atomic basins $\Omega_p$ separated by interatomic surfaces (separatrices) $S$ defined as having zero local flux in the gradient vector field $\nabla \rho_\Psi$ of the one-electron density $\rho_\Psi$ of a given wave function $\Psi$. It could also be for instance the non-atomic partition realized by the electron localization function (ELF)~\cite{BecEdg-JCP-90,SilSav-NAT-94}.

The partition ${\cal P}$ generates an orthogonal direct-sum decomposition of the one-electron Hilbert space
\begin{equation}
{\calh} = {\calh}_{\Omega_1} \oplus {\calh}_{\Omega_2} \oplus \cdots \oplus {\calh}_{\Omega_{M_{\cal P}}},
\label{}
\end{equation}
where ${\calh}_{\Omega} = L^2(\Omega \times\Sigma,\mathbb{C})$ is the one-electron Hilbert space restricted to the domain $\Omega$. It is then natural to consider the \textit{broken} spin-orbital space ${\calw}_{\cal P}$ obtained with the following direct sum
\begin{equation}
{\calw}_{\cal P} = {\calw}_{\Omega_1} \oplus {\calw}_{\Omega_2} \oplus \cdots \oplus {\calw}_{\Omega_{M_{\cal P}}},
\label{wPdecomp}
\end{equation}
where ${\calw}_{\Omega} = H^1(\Omega \times\Sigma,\mathbb{C})$ is the spin-orbital space restricted to the domain $\Omega$. The space ${\calw}_{\cal P}$ is known as the piecewise or broken first-order Sobolev space~\cite{DipErn-BOOK-12} and can also be written as
\begin{equation}
\textstyle
\calw_{\cal P} = \Big\{ \phi \in \calh \;|\; \forall \Omega \in {\cal P}, \phi_{|\Omega\times\Sigma} \in {\calw}_{\Omega} \Big\},
\label{}
\end{equation}
where $\phi_{|\Omega\times\Sigma}$ designates the restriction of $\phi$ to $\Omega\times\Sigma$. In other words, for a broken spin-orbital $\phi \in \calw_{\cal P}$, $\nabla \phi$ is only required to be square integrable on each domain of the partition but not across the domains, i.e. it can have infinite kinetic energy $\braket{\phi}{\hat{t} \; \phi}$ due to discontinuities at the interfaces between the domains. Thus, the broken spin-orbital space ${\calw}_{\cal P}$ is larger than the normal spin-orbital space ${\calw}=H^1 (\mathbb{R}^3 \times\Sigma,\mathbb{C})$ in which we would a priori want to work, i.e. $\calw \subset \calw_{\cal P}$. Yet, it would be desirable to work in the broken spin-orbital space ${\calw}_{\cal P}$ since, the direct-sum decomposition in Eq.~(\ref{wPdecomp}) would then give us a natural and rigorous definition of subspaces of strictly localized spin-orbitals, independent from any finite basis-set approximation. This is precisely what the discontinuous Galerkin method allows one to do, i.e. to work on the broken spin-orbital space ${\calw}_{\cal P}$.

The symmetric interior penalty version of the discontinuous Galerkin method~\cite{Arn-SIAM-82,ArnBreCocMar-SIAM-02,DipErn-BOOK-12} allows one to handle the decomposition in Eq.~(\ref{wPdecomp}) within a variational framework similar to that employed in quantum chemistry. To achieve this, the kinetic-energy sesquilinear form $\braket{\phi}{\hat{t}\; \psi}$ on the spin-orbital space $\calw$ in Eq.~(\ref{phitpsi}) is extended to a sesquilinear form $\braket{\phi}{\hat{t}_{\cal P}\; \psi}$ on the broken spin-orbital space $\calw_{\cal P}$. For simplicity, we will give the expression of the sesquilinear form $\braket{\phi}{\hat{t}_{\cal P}\; \psi}$ for two broken spin-orbitals $\phi$ and $\psi$ assumed to be sufficiently regular inside each domain $\Omega_p$ of the partition. It is~\cite{LinLuYinE-JCP-12,LiLin-SIAM-19}
\begin{eqnarray}
\braket{\phi}{\hat{t}_{\cal P}\; \psi} &=& \frac{1}{2} \sum_{\Omega\in{\cal P}} \int_{\Omega\times \Sigma} \nabla \phi_{|\Omega\times\Sigma}^*(\b{x}) \cdot \nabla \psi_{|\Omega\times\Sigma}(\b{x}) \d \b{x} 
\nonumber\\
&& -\frac{1}{2} \sum_{S\in{\cal S}_{\cal P}} \Biggl[ 
\int_{S\times\Sigma} \llbracket\phi\rrbracket_S^*(\b{x}) \cdot \llbrace\nabla\psi \rrbrace_S(\b{x})  \, \d s(\b{x})
+\int_{S\times\Sigma} \llbrace\nabla\phi \rrbrace_S^*(\b{x}) \cdot \llbracket\psi\rrbracket_S(\b{x}) \,  \d s(\b{x})
\Biggl]
\nonumber\\
&& + \alpha_\text{pen} \sum_{S\in{\cal S}_{\cal P}} \int_{S\times\Sigma} \llbracket\phi\rrbracket_S^*(\b{x}) \cdot \llbracket\psi\rrbracket_S(\b{x}) \, \d s(\b{x}),
\label{phitPpsi}
\end{eqnarray}
where $\alpha_\text{pen}$ is a positive real number, and, for an interface $S=\partial\Omega_1 \cap \partial\Omega_2$ between the two domains $\Omega_1$ and $\Omega_2$, we define the average of $\nabla \phi$ across $S$ as
\begin{equation}
\forall \b{x} \in S \times \Sigma,\; \llbrace\nabla \phi \rrbrace_S(\b{x}) = \frac{1}{2} \Big[ \nabla \phi_{|\Omega_1\times\Sigma}(\b{x}) + \nabla \phi_{|\Omega_2\times\Sigma}(\b{x}) \Big],
\label{average}
\end{equation}
and the jump of $\phi$ across $S$ as
\begin{eqnarray}
\forall \b{x} \in S \times \Sigma,\; \llbracket \phi\rrbracket_S(\b{x}) &=& \phi|_{\Omega_1\times\Sigma}(\b{x}) \, \b{n}_{\partial \Omega_1}(\b{r}) + \phi|_{\Omega_2\times\Sigma}(\b{x}) \,\b{n}_{\partial \Omega_2}(\b{r})
\nonumber\\
&=&  \Big[ \phi_{|\Omega_1\times\Sigma}(\b{x}) - \phi_{|\Omega_2\times\Sigma}(\b{x}) \Big] \b{n}_{\partial \Omega_1}(\b{r}),
\label{jump}
\end{eqnarray}
where $\b{n}_{\partial \Omega_i}(\b{r})$ is the outward unit normal vector to $\partial \Omega_i$ at point $\b{r}$. The operator $\hat{t}_{\cal P}$ is indeed an extension of $\hat{t}$ because, for any non-broken spin-orbitals $\phi,\psi\in {\calw}$, the jumps vanish across interfaces, i.e. $\llbracket \phi\rrbracket_S=\llbracket \psi\rrbracket_S = \b{0}$, and thus $\braket{\phi}{\hat{t}_{\cal P}\; \psi} = \braket{\phi}{\hat{t}\; \psi}$. The second term on the right-hand side of Eq.~(\ref{phitPpsi}) ensures consistency, i.e. for any spin-orbital $\psi\in H^2 (\mathbb{R}^3 \times\Sigma,\mathbb{C})$ and any broken spin-orbital $\phi$, we have $\braket{\phi}{\hat{t}_{\cal P}\; \psi} = -(1/2) \braket{\phi}{\Delta\psi}$, which can be checked by integration by parts. The third term on the right-hand side of Eq.~(\ref{phitPpsi}) ensures symmetry, i.e. $\braket{\phi}{\hat{t}_{\cal P}\; \psi} = \braket{\psi}{\hat{t}_{\cal P}\; \phi}^*$. Finally, the fourth term on the right-hand side of Eq.~(\ref{phitPpsi}) introduces penalties for jumps across the interfaces. For a sufficiently large positive value of $\alpha_\text{pen}$, the kinetic energy of any broken spin-orbital $\phi$ with non-negligible jumps at interfaces is sufficiently large for not altering the total energy minimum when $\hat{t}_{\cal P}$ is used in the energy minimization.

Starting from the broken spin-orbital space $\calw_{\cal P}$, we now define a corresponding space of admissible normalized $N$-electron broken wave functions
\begin{equation}
\textstyle
{\cal W}_{\cal P} = \Big\{ \Psi \in \bigwedge^N {\calw}_{\cal P} \, \;|\; \braket{\Psi}{\Psi}=1 \Big\} \subset {\cal H},
\label{WNP}
\end{equation}
and a $N$-electron Hamiltonian
\begin{equation}
\hat{H}_{\cal P} = \hat{T}_{\cal P} + \hat{V}_{\tne} + \hat{W}_\ee,
\label{HP}
\end{equation}
where $\hat{T}_{\cal P}=\sum_{i=1}^{N} \hat{t}_{{\cal P},i}$ is the total kinetic-energy operator corresponding to the one-electron operator $\hat{t}_{\cal P}$. The exact ground-state energy $E_0$ can then be expressed as an infimum over broken wave functions $\Psi \in {\cal W}_{\cal P}$
\begin{equation}
E_{0} = \inf_{\Psi \in {\cal W}_{\cal P}} \braket{\Psi}{\hat{H}_{\cal P} \Psi},
\label{E0NfromP}
\end{equation}
provided, we recall, that the penalty parameter $\alpha_\text{pen}$ in Eq.~(\ref{phitPpsi}) is sufficiently large for not altering the energy minimum.

Thus, the interior-penalty discontinuous Galerkin method allows us to use the broken spin-orbital space $\calw_{\cal P}$ as a variational spin-orbital space. The direct-sum decomposition of this space in Eq.~(\ref{wPdecomp}) then means that we can define orthonormal bases of strictly localized spin-orbitals. To this end, let us consider an orthonormal basis $\{\tilde{\phi}_{p,\mu} \}_{\mu \in \mathbb{N}} \subset \calw_{\Omega_p}$ of the one-electron Hilbert space ${\calh}_{\Omega_p}$ for the domain $\Omega_p$. If, for each function $\tilde{\phi}_{p,\mu}$, we introduce the function ${\phi}_{p,\mu}$ which extends it to the whole domain as
\begin{equation}
\forall \b{x} \in \mathbb{R}^3 \times\Sigma,\; {\phi}_{p,\mu}(\b{x}) = 
\begin{cases}
  \tilde{\phi}_{p,\mu}(\b{x})  & \text{if}\; \b{x} \in \Omega_p\times\Sigma\\
  0 & \text{if}\; \b{x} \not\in \Omega_p\times\Sigma
\end{cases},
\label{phipiextend}
\end{equation}
we obtain indeed a broken spin-orbital orthonormal basis $\{\phi_{p,\mu} \}^{p=1,...,M_{\cal P}}_{\mu \in \mathbb{N}} \subset \calw_{\cal P}$ of the one-electron Hilbert space $\calh$. The basis function $\phi_{p,\mu}$ corresponds to a broken spin-orbital with spatial support on the domain $\Omega_p$. Note that two spin-orbitals $\phi_{p,\mu}\in {\calw}_{\Omega_p}$ and $\phi_{p',\nu} \in {\calw}_{\Omega_{p'}}$ for two different domains ${\Omega_p}\not={\Omega_{p'}}$ are automatically orthogonal since they do not overlap. The space of admissible normalized $N$-electron broken wave functions can then be explicitly written as
\begin{eqnarray}
{\cal W}_{\cal P} &=& \Big\{ \Psi = \sum_{I \in {\cal D}_{\cal P}} C_{I} \; \phi_{I_1} \wedge \phi_{I_2} \wedge \cdots \wedge \phi_{I_N}  \;|\; \braket{\Psi}{\Psi}=1 ; (C_{I})_{I \in {\cal D}_{\cal P}} \in \mathbb{C}^{{\cal D}_{\cal P}} \Big\},
\end{eqnarray}
where $\phi_{I_1} \wedge \phi_{I_2} \wedge \cdots \wedge \phi_{I_N}$ is a Slater determinant made of $N$ localized spin-orbitals $\{\phi_{I_n}\}_{n=1,...,N}$ with the double index $I_n=(p_n,\mu_n) \in \llbracket 1, M_{\cal P}\rrbracket\times\mathbb{N}$ and ${\cal D}_{\cal P}$ is the set of ordered localized spin-orbital labels of all $N$-electron Slater determinants
\begin{equation}
{\cal D}_{\cal P} = \left\{ I=(I_1,I_2,...,I_N) \in (\llbracket 1, M_{\cal P}\rrbracket\times\mathbb{N})^N \;|\; I_1<I_2<\cdots<I_N \right\},
\label{DNP}
\end{equation}
where here ``$<$'' refers to the lexicographic order of pairs of numbers.

For any given domain $\Omega_p$, the spin-orbital basis $\{\phi_{p,\mu} \}_{\mu \in \mathbb{N}}$ for this domain is arbitrary. One possible choice is to use the eigenfunctions of the one-particle reduced density matrix for the domain $\Omega_p$, defined as follows. For a wave function $\Psi \in {\cal W}_{\cal P}$, the total one-particle reduced density matrix is
\begin{equation}
\gamma(\b{x},\b{x}') = N \int_{(\mathbb{R}^3\times\Sigma)^{N-1}} \Psi^*(\b{x}',\b{x}_2,...,\b{x}_N) \; \Psi(\b{x},\b{x}_2,...,\b{x}_N) \d \b{x}_2 ... \d \b{x}_N,
\label{}
\end{equation}
and can be expressed in the broken spin-orbital orthonormal basis $\{\phi_{p,\mu} \}^{p=1,...,M_{\cal P}}_{\mu \in \mathbb{N}}$ as
\begin{equation}
\gamma(\b{x},\b{x}') = \sum_{p=1}^{M_{\cal P}} \sum_{p'=1}^{M_{\cal P}} \sum_{\mu=1}^{\infty} \sum_{\nu=1}^{\infty} \gamma_{p,\mu;p',\nu} \phi_{p,\mu}(\b{x}) \phi_{p',\nu}^*(\b{x}'),
\label{}
\end{equation}
where $\gamma_{p,\mu;p',\nu} = \int_{(\mathbb{R}^3\times\Sigma)^{2}} \phi_{p,\mu}^*(\b{x}) \; \gamma(\b{x},\b{x}') \; \phi_{p',\nu}(\b{x}') \d\b{x}\d\b{x}'$ are the matrix elements of $\gamma$. The one-particle reduced density matrix for the domain $\Omega_p$ can then be defined as
\begin{equation}
\gamma_p(\b{x},\b{x}') = \sum_{p=1}^{M_{\cal P}} \sum_{\mu=1}^{\infty} \sum_{\nu=1}^{\infty} \gamma_{p,\mu;p,\nu} \phi_{p,\mu}(\b{x}) \phi_{p,\nu}^*(\b{x}'),
\label{}
\end{equation}
which is zero for $\b{x} \not\in \Omega_p$  or $\b{x}' \not\in \Omega_p$.
The eigenfunctions $\Big\{\phi_{p,\mu}^\text{nat} \Big\}_{\mu \in \mathbb{N}}$ of $\gamma_p(\b{x},\b{x}')$ may be called natural spin-orbitals for the domain $\Omega_p$ and are given by the eigenvalue equation
\begin{equation}
\int_{\Omega_p\times\Sigma} \gamma_p(\b{x},\b{x}') \phi_{p,\mu}^\text{nat}(\b{x}') \d \b{x}' = n_{p,\mu} \, \phi_{p,\mu}^\text{nat}(\b{x}),
\label{}
\end{equation}
where the eigenvalues $n_{p,\mu}$ are occupation numbers. For the QTAIM partition, these eigenfunctions $\Big\{\phi_{p,\mu}^\text{nat} \Big\}_{\mu \in \mathbb{N}}$ correspond to the (correlated) effective atomic orbitals~\cite{May-IJQC-14}. They could also be considered as a strictly localized version of the natural atomic orbitals of Weinhold (see, e.g., Ref.~\onlinecite{Wei-JCC-12}). Alternatively, by considering the eigenfunctions of the one-particle reduced density matrix for two adjacent domains corresponding to two bonded atoms, we could define a strictly localized version of natural bond orbitals~\onlinecite{Wei-JCC-12}.

For restricted variational wave-function spaces, as discussed in Section~\ref{sec:restrict_wf_space}, the localization constraints, i.e. imposing that $\phi_{p,\mu}$ has spatial support on the domain ${\Omega_p}$, are generally incompatible with full energy minimization. In other words, the two direct-sum decompositions in Eqs.~\eqref{wdecompCAV} and~\eqref{wPdecomp} are generally incompatible. There are then two possibilities for restricted variational wave-function spaces. The first possibility is to fully minimize the energy by releasing the localization constraints, i.e. performing a common unitary (or, more generally, invertible) transformation $\hat{U}$ mixing all the strictly localized spin-orbitals $\{\phi_{p,\mu} \}^{p=1,...,M_{\cal P}}_{\mu \in \mathbb{N}}$. This is equivalent to Eqs.~\eqref{ED} and~\eqref{WD}, but starting from the strictly localized spin-orbitals $\{\phi_{p,\mu} \}^{p=1,...,M_{\cal P}}_{\mu \in \mathbb{N}}$, which are just used as a discontinuous basis set, in the spirit of Refs.~\onlinecite{LinLuYinE-JCP-12,HuLinYan-JCP-15,ZhaLinHuYanPas-JCP-17,LiLin-SIAM-19}. The second possibility is to minimize the energy with the localization constraints, i.e. for each domain $\Omega_p$ performing a separate unitary (or, more generally, invertible) transformation $\hat{U}_p$ among the strictly localized spin-orbitals $\{\phi_{p,\mu} \}_{\mu \in \mathbb{N}}$ of this domain. In this latter case, the lowest energy is not reached, but the obtained orbitals remain strictly localized. This gives a version of the VB self-consistent-field method~\cite{LenBal-JCP-83} with well-defined strictly localized spin-orbital subspaces, even in the complete basis-set limit.

Even with restricted variational wave-function spaces, in the complete basis-set limit, the minimizing energy and wave function will be independent from the penalty constant $\alpha_\text{pen}$, provided it is large enough. In a finite basis set, the results may be depend on $\alpha_\text{pen}$, and there might be tradeoff between a too small $\alpha_\text{pen}$ which can lead to strongly discontinuous orbitals at the interfaces and a too large $\alpha_\text{pen}$ which can lead to orbitals far from being optimal away from the interfaces.

\subsection{Comparison with valence-bond theory and chemical interpretability}

For an atomic partition such as the QTAIM one, if we write a broken $N$-electron wave function, $\Psi = \sum_{I} C_{I} \; \Phi_I$, with chemically meaningful spin-singlet configurations $\{\Phi_I\}$ constructed from strictly localized spin-orbitals $\{\phi_{p,\mu}\}$, we obtain a type of classical VB expansion~\cite{ShaHib-BOOK-08} where each spin-orbital $\phi_{p,\mu}$ is spatially supported in an atomic domain $\Omega_p$.

In the classical VB approach, localized orbitals are defined in the approximate finite-dimensional setting using the structure of standard quantum-chemistry atom-centered basis sets $\{ \chi_{p,\mu} \}_{\mu=1,...,m_p}^{p=1,...,M_\text{n}} \subset {\calw}$ where $p$ runs over the $M_\text{n}$ nuclei and $m_p$ is the number of basis functions for atom $p$. These basis sets are thus naturally partitioned into $M_\text{n}$ sets of basis functions $\{ \chi_{p,\mu} \}_{\mu=1,...,m_p}$ centered on atom $p$. Typically, the basis functions $\chi_{p,\mu}$ are GTO functions (see, e.g., Ref.~\onlinecite{HelJorOls-BOOK-02}). The nonorthogonal localized orbitals $\{ \psi_{p,\mu}\}$ used in classical VB are then defined by expanding $\psi_{p,\mu}$ only on the atomic basis functions centered on atom $p$~\cite{StoWagPre-TCA-80}
\begin{equation}
\psi_{p,\mu} = \sum_{\nu=1}^{m_p} c_{p,\nu} \; \chi_{p,\nu},
\label{psipi}
\end{equation}
with coefficients $c_{p,\nu} \in \mathbb{C}$. More generally, $p$ may index a group of atoms (or fragment) of the molecule. This is known as ``strictly localized orbitals''~\cite{ShaHib-BOOK-08} or, sometimes, as ``extremely localized orbitals''~\cite{SirGenCivPieGhi-TCA-07} or ``absolutely localized orbitals''~\cite{KhaHeaBel-JCP-06} (see also Ref.~\onlinecite{MccFauZhuOgoQiuWhiRabLin-NJP-20,FauWuLin-RMS-22}). Note, however, that this corresponds to strict localization in the set of indices of basis functions, but not strict localization in real space since the atomic basis functions have infinite extent. Even though this definition of localized orbitals is convenient for practical calculations, it has the serious drawback of depending on the partition of the basis set used, which tends to become meaningless in the limit of a complete basis set. Indeed, for example, a complete basis set for any molecule can formally be obtained by a complete GTO basis set centered around a single atom, in which case there is only one set in the partition, and Eq.~(\ref{psipi}) does not impose any localization constraints.

By contrast, in the present approach, the partition of the broken spin-orbital orthonormal basis into $M_{\cal P}$ sets of spin-orbitals $\{\phi_{p,\mu} \}_{\mu \in \mathbb{N}} \subset {\calw}_{\Omega_p}$ is defined independently of an underlying basis set but directly in the infinite-dimensional setting. In other words, the subspaces of localized spin-orbitals are defined in the complete basis-set limit. Also, contrary to the classical VB approach, the present localized spin-orbitals are orthonormal and strictly localized in space, and two spin-orbitals belonging to two different domains have vanishing overlap. The VB-like structure $\Phi_I$ are thus also orthonormal and the weight of the VB-like structure $\Phi_I$ in the total wave function $\Psi$ can be unambiguously defined as $W_I = |C_I|^2$ (the so-called Chirgwin-Coulson, L\"owdin, and inverse weights of the usual VB approach~\cite{ShaHib-BOOK-08} all reduce to this formula for orthonormal VB structures).

Another nice interpretative feature of the strictly localized spin-orbitals $\{\phi_{p,\mu} \}^{p=1,...,M_{\cal P}}_{\mu \in \mathbb{N}}$ is that the electron population on the domain $\Omega_p$ is simply given by 
\begin{equation}
N_p = \int_{\Omega_p\times\Sigma} \gamma(\b{x},\b{x}) \d\b{x} = \sum_{\mu=1}^{\infty} \gamma_{p,\mu;p,\mu}.
\label{}
\end{equation}
For the QTAIM partition, $N_p$ is the QTAIM electron population on atom $p$. Interestingly, owing to the orthonormality and strict localization of the spin-orbital basis, it also corresponds to the Mulliken~\cite{Mul-JCP-55} and Löwdin~\cite{Low-JCP-50} electron populations, which are identical for an orthonormal basis. This point was already pointed out in Refs.~\onlinecite{RamSalMay-JCP-13,May-IJQC-14}.

\subsection{Practical construction of finite strictly localized spin-orbital basis sets}

In practice, we need to choose finite strictly localized spin-orbital orthonormal bases $\{\tilde{\phi}_{p,\mu} \}_{\mu=1,..,m_p} \subset \calw_{\Omega_p}$ for each domain $\Omega_p$. When $\Omega_p$ is an atomic domain, the simplest possibility is to take a standard quantum-chemistry atom-centered basis set $\{ \chi_{p,\mu} \}_{\mu=1,...,m_p}^{p=1,...,M_\text{n}}$ (for example, a GTO basis set) and restrict the basis functions $\chi_{p,\mu}$ centered around the nucleus $p$ to the domain $\Omega_p$
\begin{equation}
\tilde{\chi}_{p,\mu} = \chi_{p,\mu\;|_{\Omega_p\times\Sigma}}.
\label{}
\end{equation}
These truncated basis functions may then be orthonormalized, e.g. using L\"owdin symmetric orthonormalization~\cite{Low-JCP-50},
\begin{equation}
\tilde{\phi}_{p,\mu} = \sum_{\nu=1}^{m_p} [S_p^{-1/2}]_{\mu,\nu} \; \tilde{\chi}_{p,\nu},
\label{tildephipi}
\end{equation}
where $[S_p]_{\mu,\nu} = \braket{\tilde{\chi}_{p,\mu}}{\tilde{\chi}_{p,\nu}}=\int_{\Omega_p\times\Sigma} \tilde{\chi}_{p,\mu}^*(\b{x})\, \tilde{\chi}_{p,\nu}(\b{x}) \, \d\b{x}$ are the matrix elements of the overlap matrix of the truncated basis functions of domain $\Omega_p$. Extension of the basis functions $\tilde{\phi}_{p,\mu}$ to the whole domain according to Eq.~(\ref{phipiextend}) then leads to a finite strictly localized atomic spin-orbital orthonormal basis $\{\phi_{p,\mu} \}_{\mu=1,..,m_p}^{p=1,...,M_\text{n}} \subset \calw_{\cal P}$.

A more sophisticated possibility proposed in Refs.~\onlinecite{LinLuYinE-JCP-12,HuLinYan-JCP-15,ZhaLinHuYanPas-JCP-17} in the context of periodic KS calculations is to generate the basis set on-the-fly in each system. For each domain $\Omega_p$, the basis functions $\{ \chi_{p,\mu} \}_{\mu=1,...,m_p}$ are calculated as the $m_p$ first eigenfunctions of a one-electron (effective) Hamiltonian $\hat{h}$ (such as the HF or KS Hamiltonian)
\begin{equation}
\hat{h} \; \chi_{p,\mu} = \varepsilon_{p,\mu} \; \chi_{p,\mu},
\label{}
\end{equation}
on a domain $\Omega \supset \Omega_p$ with some boundary conditions on $\partial \Omega$. The basis functions $\{ \chi_{p,\mu} \}_{\mu=1,...,m_p}$ are then restricted to $\Omega_p$ and orthonormalized (and possible near-linear dependencies are eliminated). The resulting basis set has been called adaptive local basis set. In Ref.~\onlinecite{LiLin-SIAM-19} a variant of this approach was also proposed which constructs the local basis functions using the Hamiltonian $\hat{h}$ on the whole domain and not a restricted domain $\Omega$. For a QTAIM domain $\Omega_p$, this last variant is in fact equivalent to the procedure proposed by Mayer to obtain (uncorrelated) effective atomic orbitals~\cite{May-CJP-96,May-IJQC-14}.

Whatever the practical approach used, we would like to emphasize that the general strategy of constructing a basis set for the global Hilbert space ${\calh}$ by patching together local basis sets for the local Hilbert subspaces ${\calh}_{\Omega_p}$ solves the overcompleteness problem~\cite{KinStaNew-CPL-75} of the multicenter basis sets used in quantum chemistry. In the limit where the local basis sets are complete for the local Hilbert subspaces ${\calh}_{\Omega_p}$, their union properly forms a complete basis set for the global Hilbert space ${\calh}$.

When $\Omega_p$ is an atomic domain, the localized spin-orbital basis sets bear some resemblance with some compactly supported basis sets, namely the numerical atomic orbitals~\cite{JunPazSanArt-PRB-01} or numeric atom-centered orbitals~\cite{BluGehHanHavHavRenReuSch-CPC-09} which consist in atomic orbitals calculated with a radial confining potential such as the orbitals are strictly zero beyond a spherical cutoff radius. However, the latter approach is not based on a partition of space and thus it is not clear how to formally reach the complete basis-set limit.

We give now the expressions of the one- and two-electron integrals in a truncated non-orthogonal basis set $\{ \tilde{\chi}_{p,\mu} \}_{\mu=1,..,m_p}^{p=1,...,M_{\cal P}}$ where $\tilde{\chi}_{p,\mu}$ has support in the domain $\Omega_p$. For two functions $\tilde{\chi}_{p,\mu}$ and $\tilde{\chi}_{p',\nu}$ either on the same domain ($p=p'$) or on two different domains ($p \not= p'$) sharing an interface ($\partial \Omega_p\cap \partial \Omega_{p'} \not=\emptyset$), the kinetic-energy integrals are
\begin{eqnarray}
\braket{\tilde{\chi}_{p,\mu}}{\hat{t}_{\cal P}\; \tilde{\chi}_{p',\nu}} &=& 
\frac{1}{2} \delta_{p,p'} \int_{\Omega_p\times \Sigma} \nabla \tilde{\chi}_{p,\mu}^*(\b{x}) \cdot \nabla \tilde{\chi}_{p,\nu}(\b{x}) \, \d \b{x}
\nonumber\\
&&-\frac{1}{4} \Biggl[ 
\int_{(\partial \Omega_p\cap \partial \Omega_{p'})\times\Sigma} \tilde{\chi}_{p,\mu}^*(\b{x}) \, \b{n}_{\partial \Omega_p}(\b{r}) \cdot \nabla\tilde{\chi}_{p',\nu}(\b{x})  \, \d s(\b{x})
\nonumber\\
&&+\int_{(\partial \Omega_p\cap \partial \Omega_{p'})\times\Sigma} \nabla \tilde{\chi}_{p,\mu}^*(\b{x}) \cdot \b{n}_{\partial \Omega_{p'}}(\b{r}) \, \tilde{\chi}_{p',\nu}(\b{x})  \, \d s(\b{x}) 
\Biggl]
\nonumber\\
&& + \alpha_\text{pen} \int_{(\partial \Omega_p\cap \partial \Omega_{p'})\times\Sigma} \tilde{\chi}_{p,\mu}^*(\b{x}) \, \b{n}_{\partial \Omega_p}(\b{r}) \cdot \b{n}_{\partial \Omega_{p'}}(\b{r}) \, \tilde{\chi}_{p',\nu}(\b{x}) \, \d s(\b{x}).
\label{tpint}
\end{eqnarray}
For two functions $\tilde{\chi}_{p,\mu}$ and $\tilde{\chi}_{p,\nu}$ on the same domain $\Omega_p$, the nuclei-electron integrals are
\begin{eqnarray}
\braket{\tilde{\chi}_{p,\mu}}{\hat{v}_\tne\; \tilde{\chi}_{p,\nu}} &=& \int_{\Omega_p\times \Sigma} \tilde{\chi}_{p,\mu}^*(\b{x}) \, v_\tne(\b{r}) \, \tilde{\chi}_{p,\nu}(\b{x}) \, \d \b{x}.
\label{vneint}
\end{eqnarray}
Finally, for two functions $\tilde{\chi}_{p,\mu}$ and $\tilde{\chi}_{p',\nu}$ either on the same domain ($\Omega_p=\Omega_{p'}$) or on different domains ($\Omega_p\not=\Omega_{p'}$), the two-electron integrals are
\begin{eqnarray}
\braket{\tilde{\chi}_{p,\mu} \, \tilde{\chi}_{p',\nu}}{\hat{w}_\ee\; \tilde{\chi}_{p,\lambda} \, \tilde{\chi}_{p',\sigma}} &=& \int_{(\Omega_p\times \Sigma)\times(\Omega_{p'}\times \Sigma)} \frac{\tilde{\chi}_{p,\mu}^*(\b{x}_1) \, \tilde{\chi}_{p',\nu}^*(\b{x}_2) \, \tilde{\chi}_{p,\lambda}(\b{x}_1) \, \tilde{\chi}_{p',\sigma}(\b{x}_2)}{|\b{r}_1 -\b{r}_2|} \d \b{x}_1 \d \b{x}_2.
\label{twoeint}
\end{eqnarray}
These are the only non-zero integrals. Note in particular that the kinetic-energy integrals in Eq.~(\ref{tpint}) only involve at most two adjacent domains, which, for an atomic partition, is reminiscent of the nearest-neighbor hopping in the H\"uckel or tight-binding model (see, e.g., Ref.~\onlinecite{Pow-INC-11}). Thanks to the strict locality of the basis functions, the number of two-electron integrals scales as $O(M_{\cal P}^2)$, without having to apply any approximation. For the QTAIM partition, algorithms have already been proposed~\cite{SalDurMay-JCP-01,MarBlaFra-JCP-04} to calculate the volumic integrals over atomic basins appearing in Eqs.~(\ref{tpint})-~(\ref{twoeint}). The problem of calculating surface integrals over interatomic surfaces, similar to the ones appearing in Eq.~(\ref{tpint}), has also been addressed~\cite{Pop-TCA-01,AniAna-JCC-20}.

\section{Numerical illustrations}
\label{sec:numerical}

In this section, we illustrate the use of strictly localized orbitals based on a spatial partitioning and the interior-penalty discontinuous Galerkin method on 1D model systems. We performed the calculations reported in this section using an in-house code publicly available in Ref.~\onlinecite{FenTou-PROG-26}.

\subsection{One-electron diatomic molecules}

We first consider a 1D model of a one-electron diatomic molecule. On the spinfree Hilbert space $\calh = L^2(\mathbb{R},\mathbb{C})$, the one-electron Hamiltonian is
\begin{eqnarray}
\hat{h} = -\frac{1}{2} \frac{\d^2}{\d x^2} + v_{\tne}(x)
\label{h1e}
\end{eqnarray}
where $v_{\tne}(x)$ is the nuclei-electron potential
\begin{eqnarray}
v_\tne(x)=-Z_\L v(x-R_\L) -Z_\R v(x-R_\R),
\label{vneh2+}
\end{eqnarray}
corresponding to a left (L) nucleus and a right (R) nucleus with charges $Z_\L$ and $Z_\R$ at positions $R_\L=-R/2$ and $R_\R=R/2$. For $v$, we use the soft-Coulomb model potential~\cite{WagStoBurWhi-PCCP-12,Li-JMC-22}
\begin{eqnarray}
v(x)=\frac{1}{\sqrt{x^2+a^2}},
\label{vSC}
\end{eqnarray}
with regularization parameter $a=0.2$.
We want to find the lowest-energy molecular orbitals $\psi_i$ and orbital energies $\varepsilon_i$
\begin{eqnarray}
\hat{h} \psi_i = \varepsilon_i \psi_i.
\label{}
\end{eqnarray}
We will compare three different types of calculations: a reference grid calculation, a standard calculation in a Hermite-Gaussian basis set, and a discontinuous-Galerkin calculation in a truncated Hermite-Gaussian basis set.

\vspace{0.3cm}
\noindent
\textit{Reference grid calculation.} 
To obtain a nearly-exact reference, we use a spatial grid discretization with homogeneous Dirichlet boundary conditions, leading to the finite-difference grid Hamiltonian
\begin{eqnarray}
h^{\rm grid}_{k,k'}= - \frac{1}{2} \frac{\delta_{k,k'+1}+\delta_{k,k'-1}-2\delta_{k,k'}}{\Delta x^2}+ v_\tne(x_k)\delta_{k,k'},
\end{eqnarray}
on an interval $[-L,L]$ with $L=6.0$ and grid spacing $\Delta x=0.015$. The grid Hamiltonian is then diagonalized to obtain the reference molecular orbitals $\psi_i$ on the grid and the reference orbital energies $\varepsilon_i$.

\vspace{0.3cm}
\noindent
\textit{Standard calculations in a Hermite-Gaussian basis set.}
We use a one-electron basis set $\{ \chi_{p,n} \}_{n=0,...,n_\text{max}}^{p =\L,\R}$ of normalized Hermite-Gaussian functions centered on the two nuclei, 
\begin{equation}
\chi_{p,n}(x)=N_n H_n\!\left(\sqrt{2\alpha}(x-R_p)\right)\exp[-\alpha(x-R_p)^2],
\end{equation}
where $n \geq 0$ is an integer quantum number, $H_n$ are the Hermite polynomials, $N_n=\Big(\sqrt{2\alpha}/(2^n n!\sqrt{\pi})\Big)^{1/2}$ is the normalization factor, and $\alpha >0$ is a real constant. For a single nucleus $p$, it is well known that the set $\{ \chi_{p,n} \}_{n=0, ..., \infty}$ is a complete orthonormal basis of $L^2(\mathbb{R},\mathbb{C})$ for any fixed exponent $\alpha$. However, for two nuclei, $\{ \chi_{p,n} \}_{n=0, ..., \infty}^{p=\L,\R}$ is an overcomplete basis. This is known as a frame in functional analysis~\cite{Chr-BOOK-03,Hei-BOOK-11}. Unless indicated otherwise, we use a maximal quantum number $n_\text{max} = 10$, and the same exponent $\alpha = 1.5$ in all basis functions, instead of multiple exponents in order to avoid optimizing them. Except for that, this basis set is quite similar to the GTO basis sets widely used in quantum chemistry.

To calculate the orbitals and orbital energies, we first construct a global orthonormal basis $\{ \phi_\mu \}_{\mu=1,...,M}$ (where $M=2 (n_\text{max}+1)$) by L\"owdin symmetric orthonormalization of $\{ \chi_{p,n} \}_{n=0,...,n_\text{max}}^{p =\L,\R}$, and then diagonalize the Hamiltonian in this orthonormal basis. This leads to the molecular orbitals as linear combinations of the basis functions $\phi_\mu$,
\begin{equation}
\psi_i(x) = \sum_{\mu=1}^M c_{\mu, i} \, \phi_{\mu}(x).
\end{equation}

\begin{figure}[t]
    \centering
    \includegraphics[width=0.8\linewidth]{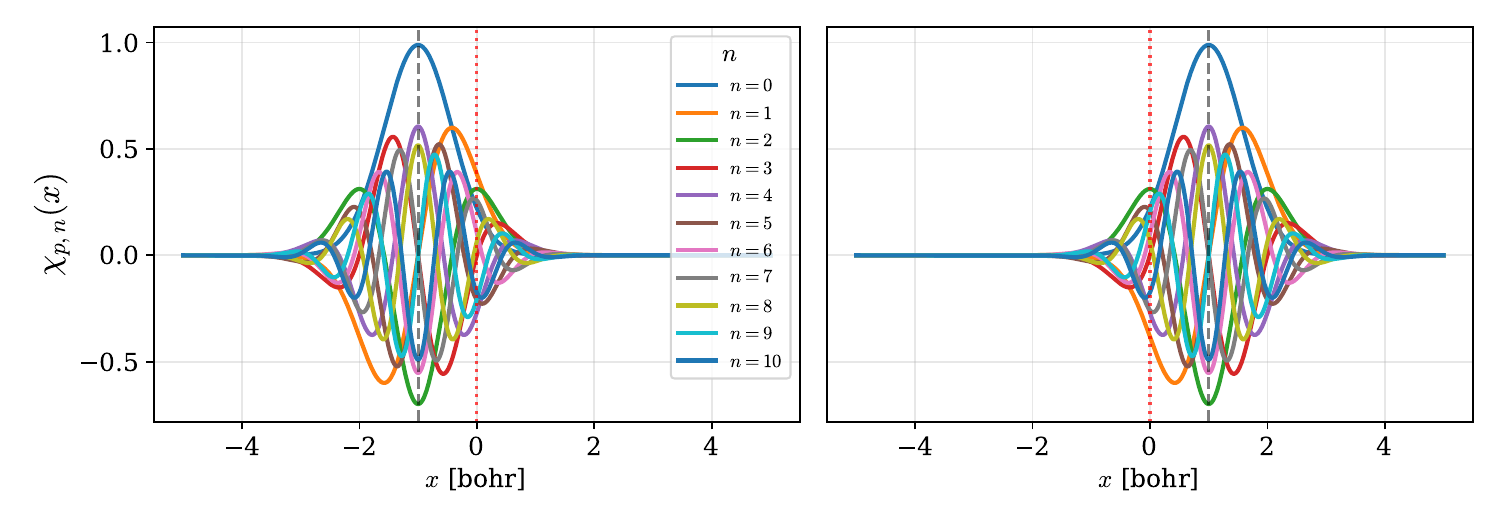}\\
    \includegraphics[width=0.8\linewidth]{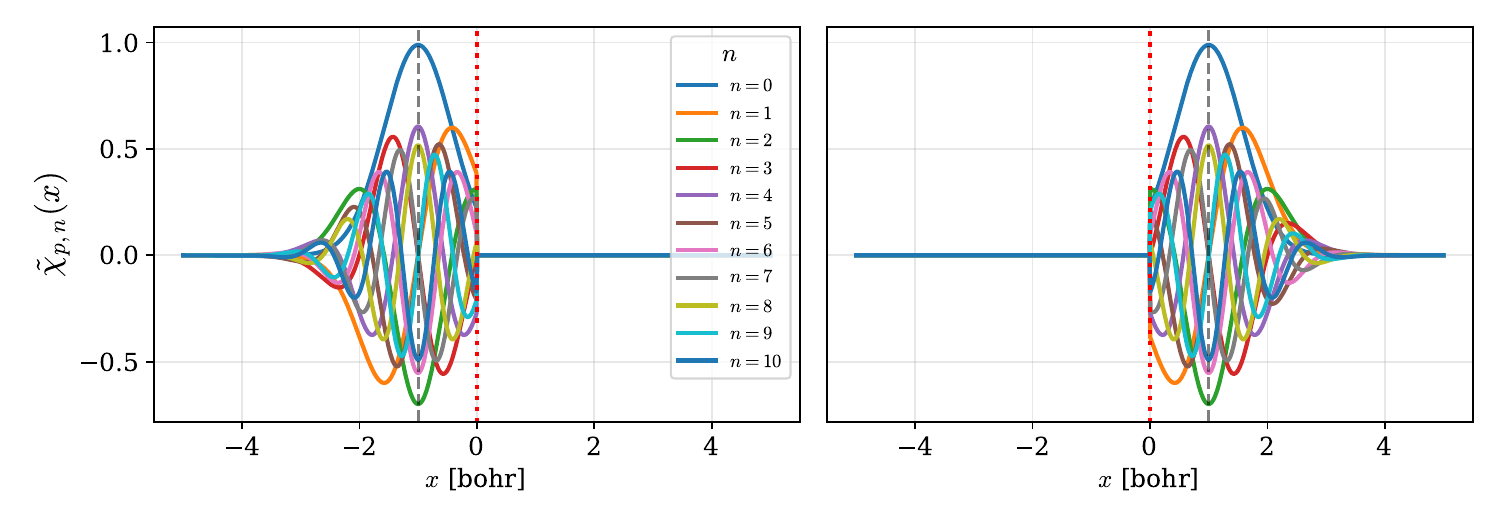}\\
    \includegraphics[width=0.8\linewidth]{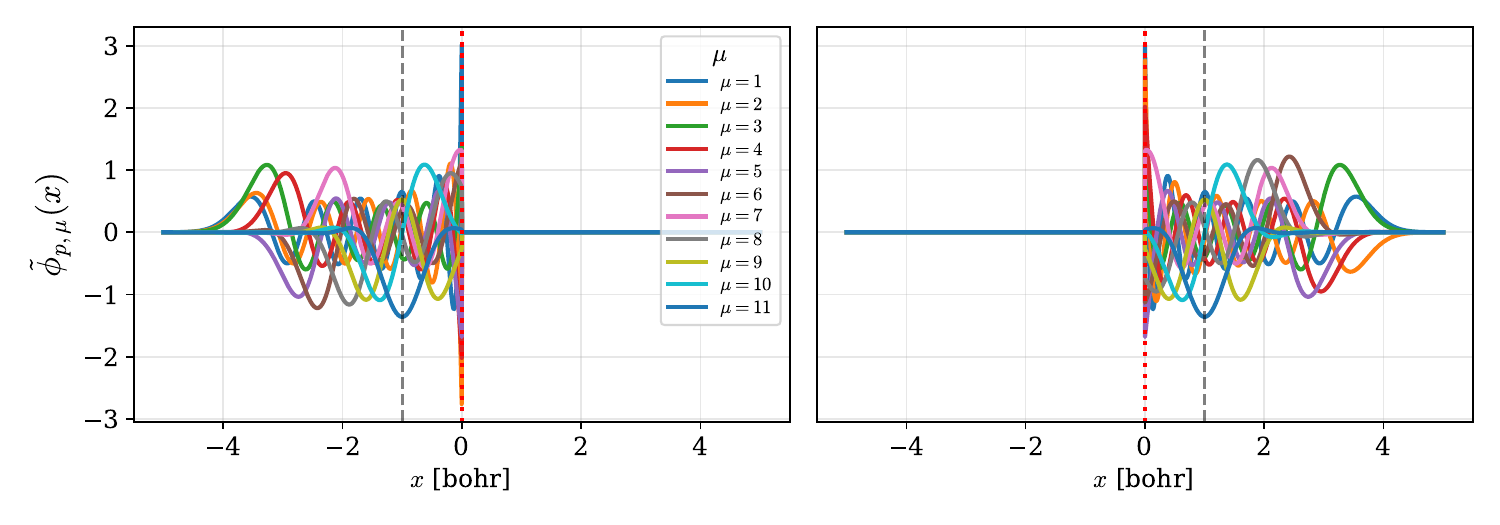}
    \caption{Different one-electron basis sets used in this work for two nuclei. First row: Hermite-Gaussian basis set $\{ \chi_{p,n} \}_{n=0,...,n_\text{max}}^{p =\L,\R}$. Second row: Truncated Hermite-Gaussian basis set $\{ \tilde{\chi}_{p,n} \}_{n=0,...,n_\text{max}}^{p =\L,\R}$. Third row: Orthonormalized truncated Hermite-Gaussian basis set $\{ \tilde{\phi}_{p,\mu} \}_{\mu=1,...,m}^{p=\L,\R}$.}
    \label{fig:basisset}
\end{figure}

\vspace{0.3cm}
\noindent
\textit{Discontinuous Galerkin calculations in a truncated Hermite-Gaussian basis set.} For simplicity, independently from the nuclear charges $Z_\L$ and $Z_\R$, we always partition the real line into two symmetric atomic domains:
\begin{equation}
\mathcal P=\{\Omega_\L,\Omega_\R\},\qquad \Omega_\L=(-\infty,0), \qquad \Omega_\R=(0,+\infty).
\end{equation}
The Hermite-Gaussian basis functions centered on each nucleus $p$ are restricted to the corresponding domain, $\tilde{\chi}_{p,n} = \chi_{p,n|_{\Omega_p}}$, and we construct a strictly localized orthonormal basis $\{ \tilde{\phi}_{p,\mu} \}_{\mu=1,...,m}^{p=\L,\R}$ (with $m = n_\text{max}+1$) by separate L\"owdin symmetric orthonormalization for each domain $\Omega_p$, as in Eq.~\eqref{tildephipi}. The original Hermite-Gaussian basis set $\{ \chi_{p,n} \}_{n=0,...,n_\text{max}}^{p =\L,\R}$, the truncated Hermite-Gaussian basis set $\{ \tilde{\chi}_{p,n} \}_{n=0,...,n_\text{max}}^{p =\L,\R}$ and the orthonormalized truncated Hermite-Gaussian basis set $\{ \tilde{\phi}_{p,\mu} \}_{\mu=1,...,m}^{p=\L,\R}$ are represented in Fig.~\ref{fig:basisset} for an internuclear distance of $R=2$.

Since the only interface between the two domains is the origin point $x=0$, i.e.  $S=\partial\Omega_\L\cap\partial\Omega_\R=\{0\}$, the kinetic-energy integrals of the interior-penalty discontinuous Galerkin method in Eq.~(\ref{tpint}) in the strictly localized orthonormal basis $\{ \tilde{\phi}_{p,\mu} \}_{\mu=1,...,m}^{p=\L,\R}$ are
\begin{eqnarray}
\braket{\tilde{\phi}_{p,\mu}}{\hat{t}_{\cal P}\,\tilde{\phi}_{p',\nu}}
&=&
\frac12\delta_{p,p'}\int_{\Omega_p}\tilde{\phi}_{p,\mu}'(x)\tilde{\phi}_{p',\nu}'(x)\,\d x
-\frac14\Bigl[\tilde{\phi}_{p,\mu}(0)\,n_{\partial\Omega_p}\,\tilde{\phi}_{p',\nu}'(0)+\tilde{\phi}_{p,\mu}'(0)\,n_{\partial\Omega_{p'}}\,\tilde{\phi}_{p',\nu}(0)\Bigr] \nonumber\\
\label{tPint1D}
&&+\alpha_\text{pen}\,\tilde{\phi}_{p,\mu}(0)\,n_{\partial\Omega_p}n_{\partial\Omega_{p'}}\,\tilde{\phi}_{p',\nu}(0),
\end{eqnarray}
where $n_{\partial\Omega_\L}=+1$ and $n_{\partial\Omega_\R}=-1$. Using these kinetic-energy integrals and the nuclei-electron integrals as in Eq.~\eqref{vneint}, we then form the Hamiltonian matrix and diagonalize it in order to calculate the orbitals and orbital energies. The molecular orbitals are thus expressed as linear combinations of the basis functions $\tilde{\phi}_{p,\mu}$,
\begin{equation}
\psi_i(x) = \sum_{p=\L,\R} \sum_{\mu=1}^{m} \tilde{c}_{p,\mu; i} \, \tilde{\phi}_{p,\mu}(x).
\end{equation}

\begin{figure}[t]
\centering
\begin{tabular}{cc}
\includegraphics[width=0.4\linewidth]{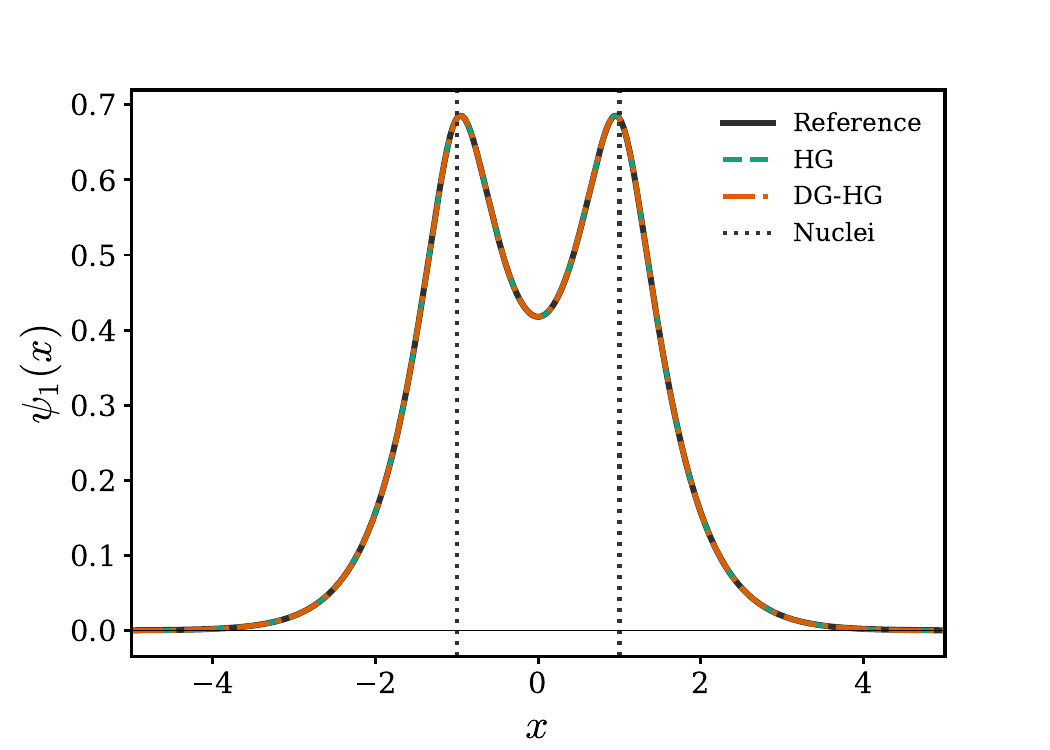} &
\includegraphics[width=0.4\linewidth]{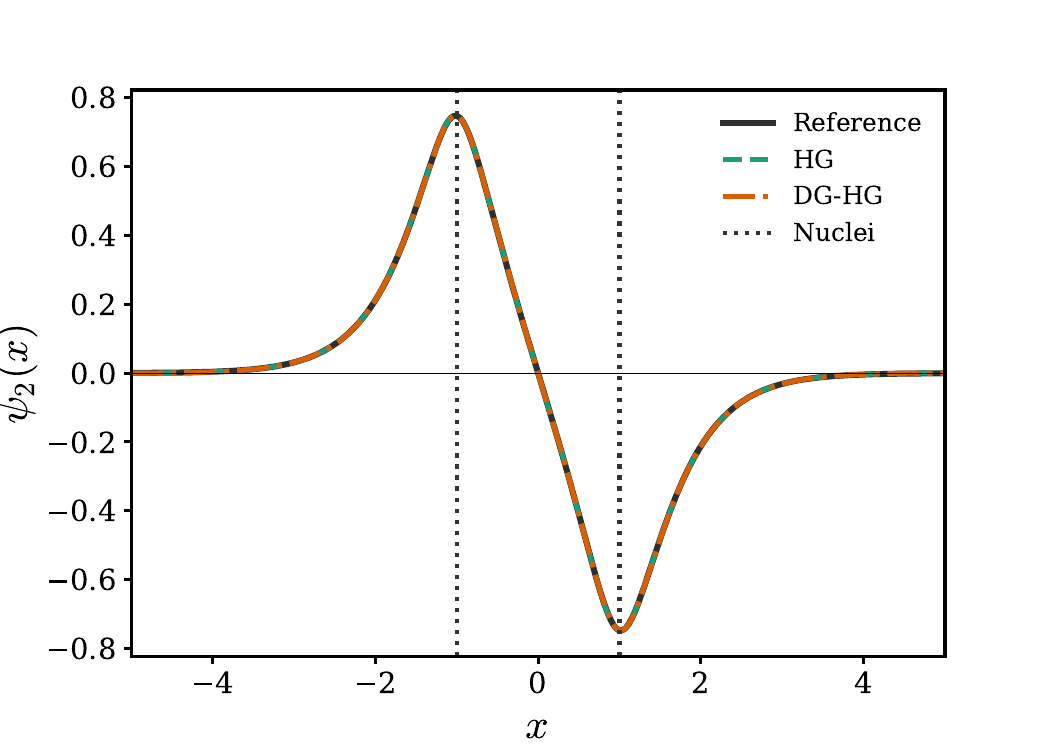} \\[0.4em]
(a) H$_2^+$ molecule, $\psi_1$ &
(b) H$_2^+$ molecule, $\psi_2$ \\[0.8em]
\includegraphics[width=0.4\linewidth]{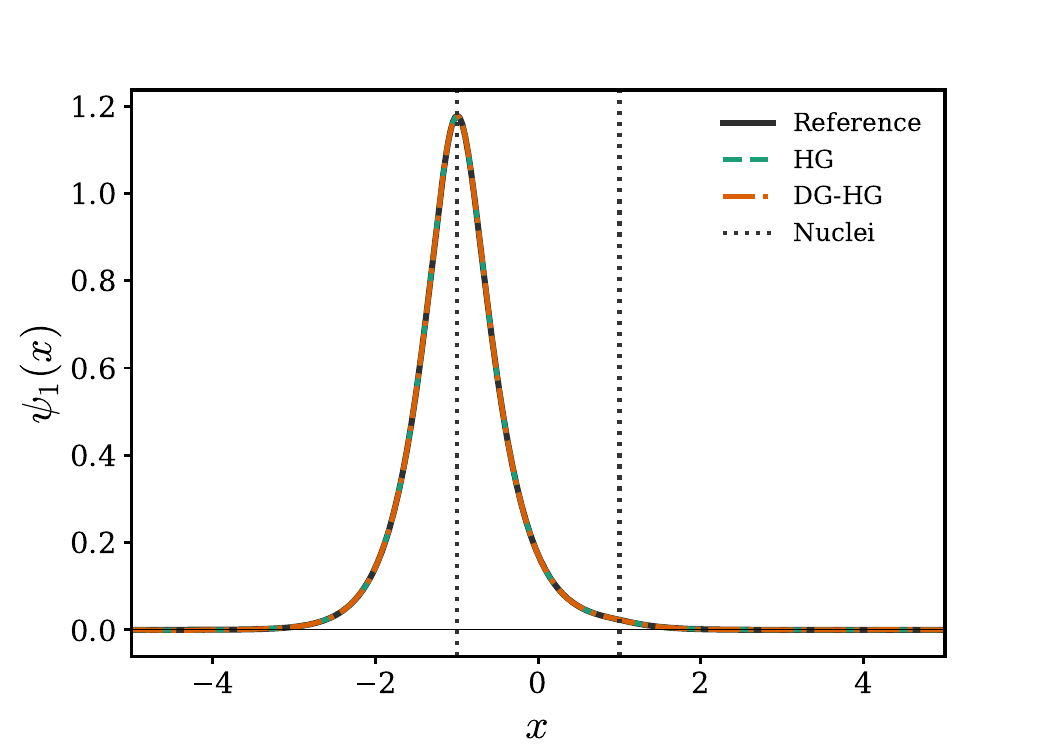} &
\includegraphics[width=0.4\linewidth]{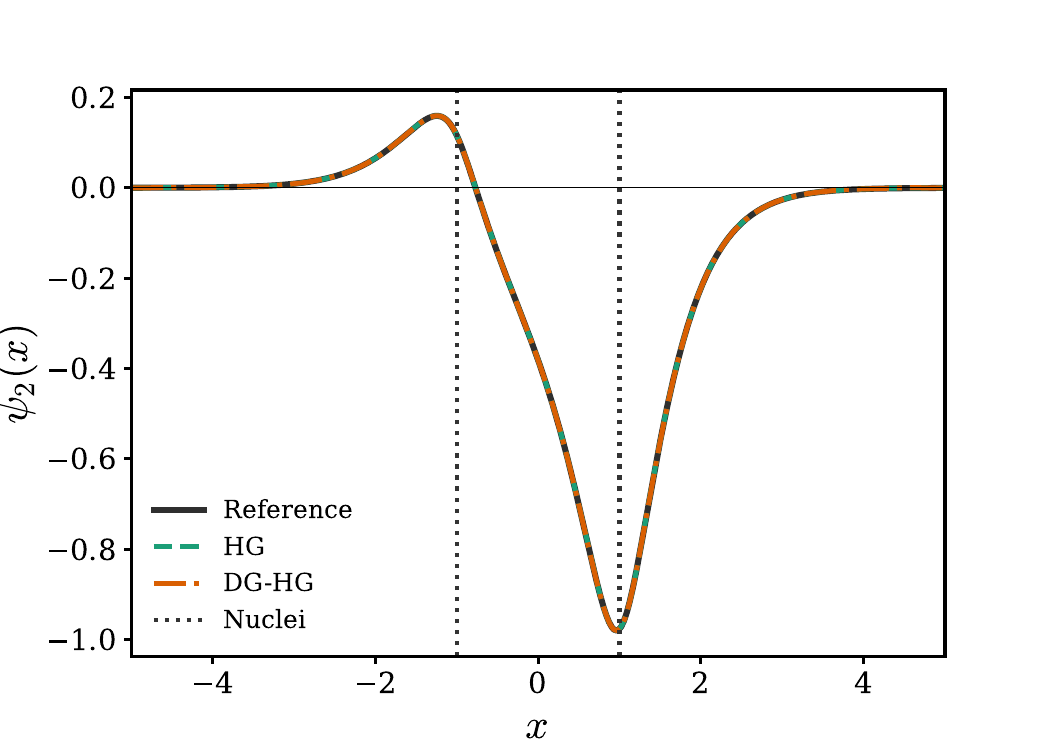} \\[0.4em]
(c) HeH$^{2+}$ molecule, $\psi_1$ &
(d) HeH$^{2+}$ molecule, $\psi_2$ \\[0.8em]
\end{tabular}
\caption{Lowest-energy molecular orbitals $\psi_1$ and $\psi_2$ of the one-dimensional H$_2^+$ and HeH$^{2+}$ molecules obtained with the reference grid method, the standard Hermite-Gaussian (HG) basis set, and the discontinuous-Galerkin Hermite-Gaussian (DG-HG) basis set.}
\label{fig:orbital_comparison}
\end{figure}

\begin{table}[t]
\caption{Energies of the two lowest-energy molecular orbitals of the one-dimensional H$_2^+$ and HeH$^{2+}$ molecules obtained with the reference grid method, the standard Hermite-Gaussian (HG) basis set, and the discontinuous-Galerkin Hermite-Gaussian (DG-HG) basis set.}
\centering
\begin{tabular}{llcc}
\hline
\hline
System & Method & $\varepsilon_1$ & $\varepsilon_2$ \\
\hline
{H$_2^+$}
&Reference & -3.039651 & -2.698105 \\
&HG        & -3.038777 & -2.696328 \\
&DG-HG     & -3.038658 & -2.696859 \\
\hline
{HeH$^{2+}$}
&Reference & -6.119287 & -3.430775 \\
&HG        & -6.112563 & -3.429280 \\
&DG-HG     & -6.113543 & -3.429679 \\
\hline
\hline
\end{tabular}
\label{tab:onebody_energies}
\end{table}

\vspace{0.3cm}
\noindent
\textit{Results.} We consider the case of nuclear charges $Z_\L=Z_\R=1$, corresponding to the one-dimensional H$_2^+$ molecule, and the case of nuclear charges $Z_\L=2$ and $Z_\R=1$, corresponding to the one-dimensional HeH$^{2+}$ molecule, both at a fixed internuclear distance of $R=2$. Figure~\ref{fig:orbital_comparison} compares the two lowest-energy molecular orbitals $\psi_1$ and $\psi_2$ obtained with the reference method, the standard Hermite-Gaussian basis set, and the discontinuous Galerkin Hermite-Gaussian basis set, while Table~\ref{tab:onebody_energies} reports the corresponding orbital energies. The standard Hermite-Gaussian basis set and discontinuous Galerkin Hermite-Gaussian basis set produce nearly indistinguishable molecular orbitals. The two methods also give very similar orbital energies, in good agreement with the reference.

\begin{figure}[t]
\centering
\begin{tabular}{cc}
\includegraphics[width=0.4\linewidth]{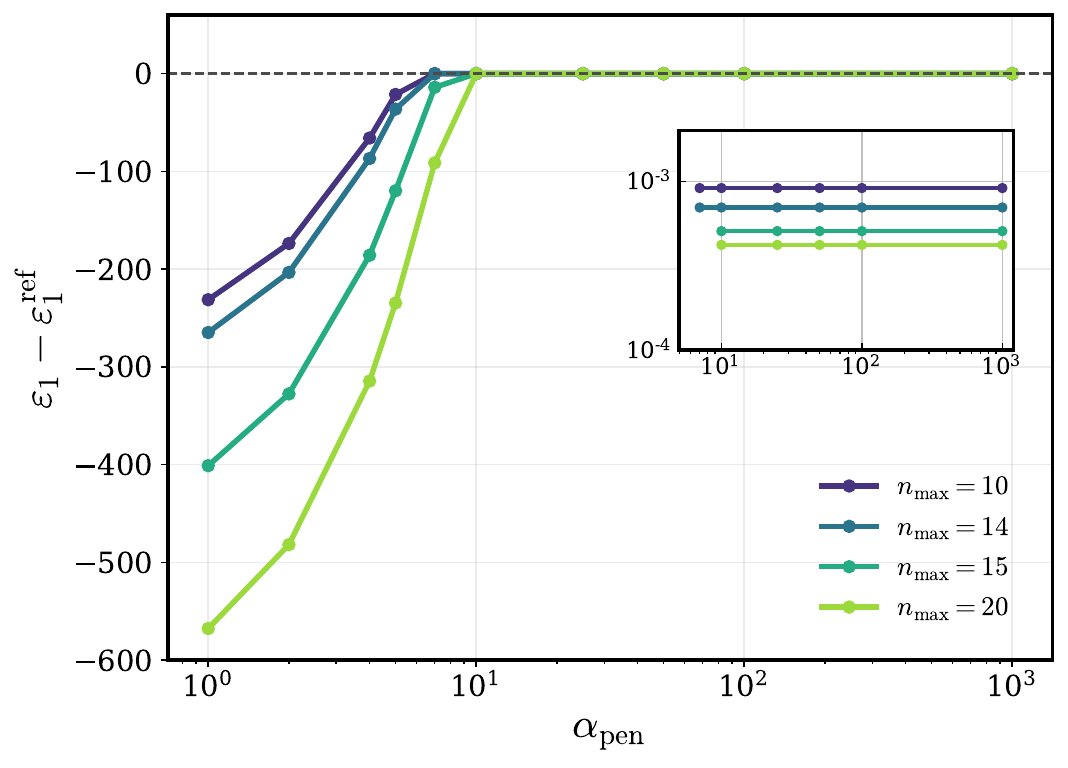} &
\includegraphics[width=0.4\linewidth]{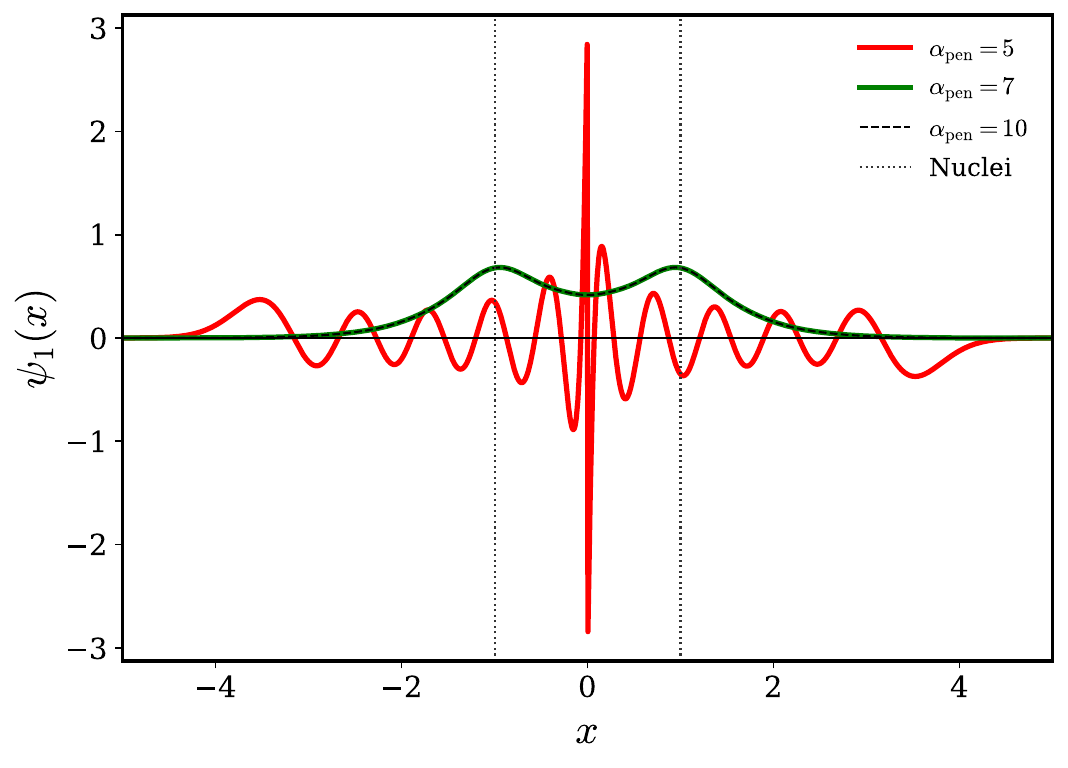} \\[0.4em]
(a) H$_2^+$ molecule, error on lowest orbital energy $\varepsilon_1$ &
(b) H$_2^+$ molecule, lowest molecular orbital $\psi_1$ \\[0.8em]
\includegraphics[width=0.4\linewidth]{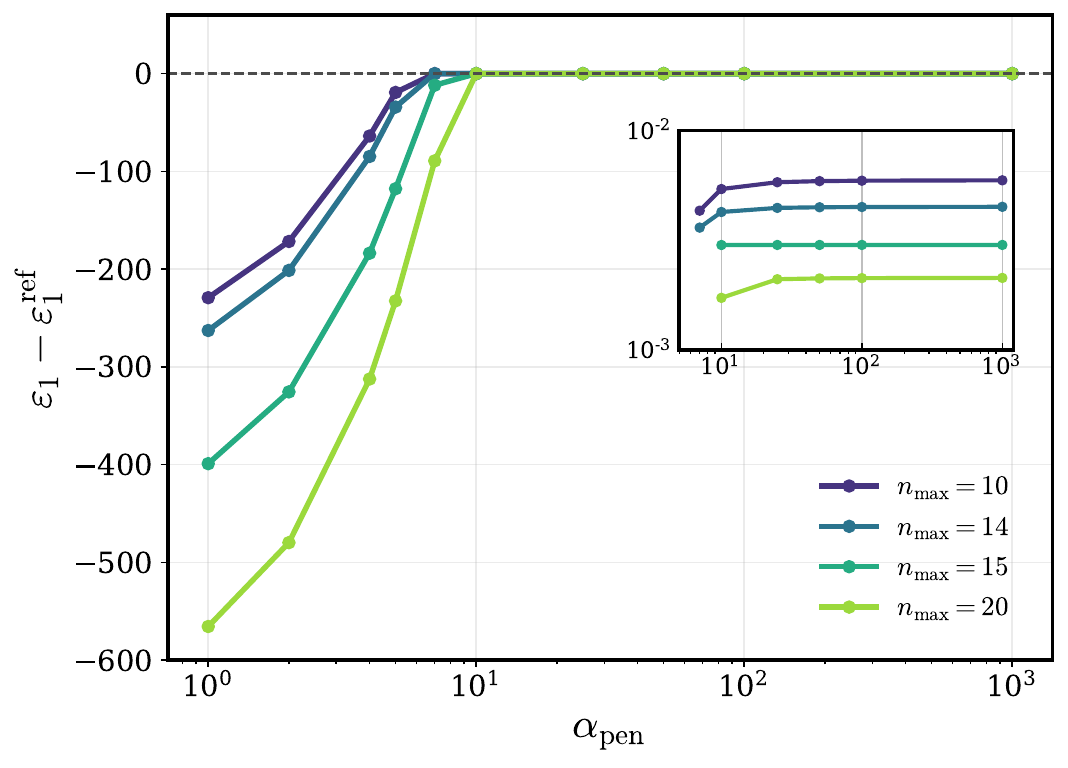} &
\includegraphics[width=0.4\linewidth]{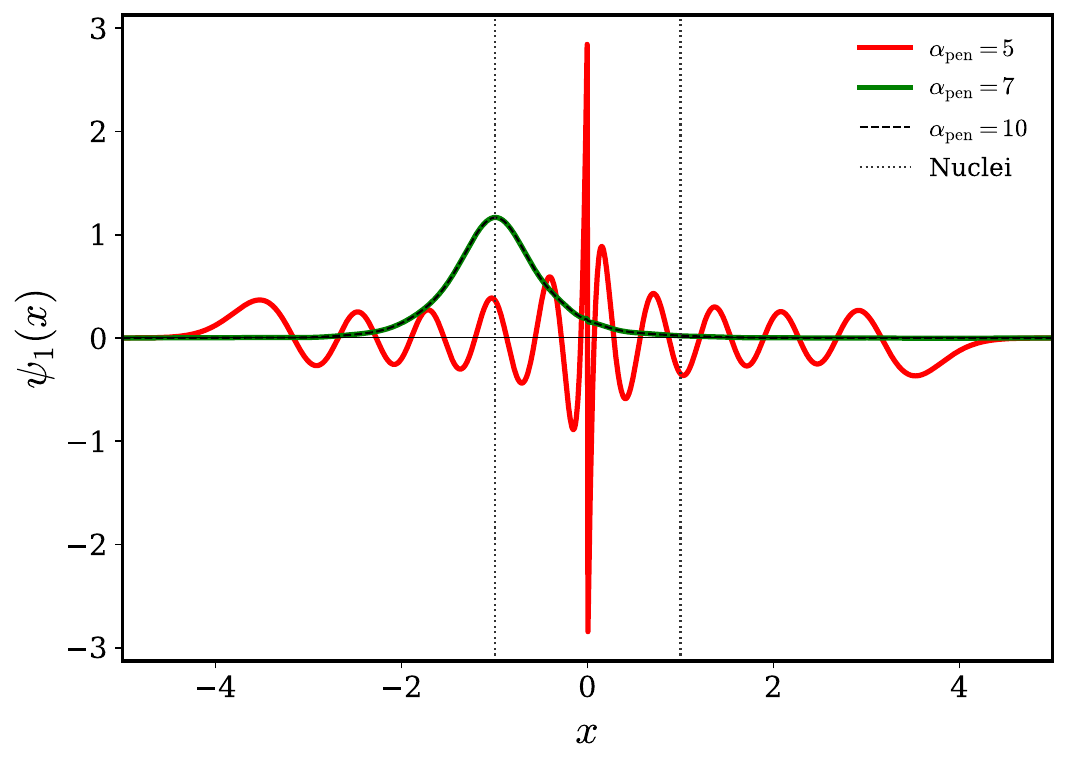} \\[0.4em]
(c) HeH$^{2+}$ molecule, error on lowest orbital energy $\varepsilon_1$ &
(d) HeH$^{2+}$ molecule, lowest molecular orbital $\psi_1$ \\[0.8em]
\end{tabular}
\caption{Dependence of the discontinuous-Galerkin calculations on the penalty parameter $\alpha_{\rm pen}$ for the one-dimensional H$_2+$ and  HeH$^{2+}$ molecules. Left panel reports the error on the lowest orbital energy $\varepsilon_1$ as a function of $\alpha_{\rm pen}$ for several basis sizes $n_\text{max}$. Right panel compares the corresponding lowest molecular orbital $\psi_1$ for different values of $\alpha_{\rm pen}$.}
\label{fig:alpha_penalty}
\end{figure}

In Figure~\ref{fig:alpha_penalty} we explore the dependence of the discontinuous-Galerkin calculations on the penalty parameter $\alpha_{\rm pen}$. The error on the lowest orbital energy $\varepsilon_1$ highlights the existence of a threshold value of the order of 10 for the penalty parameter. Above this threshold, the orbital energy becomes only very weakly dependent on $\alpha_{\rm pen}$ and the continuity across the interface is successfully imposed in the orbital. As the basis size $n_\text{max}$ increases, the location of this threshold remains nearly unchanged. In contrast, for $\alpha_{\rm pen}$ below the threshold, the discontinuous-Galerkin solution deviates significantly from the reference calculation. The lowest molecular orbital develops a pronounced discontinuity at the interface and the corresponding orbital energy becomes artificially too low. This indicates that the continuity constraint is insufficiently enforced, allowing the variational procedure to lower the energy through unphysical discontinuous solutions, as illustrated by the red curve in Fig.~\ref{fig:alpha_penalty}(b) or (d). For the remaining calculations, we set the penalty parameter to $\alpha_{\rm pen}=15$, which provides a safety margin above the observed threshold. The dashed black curves in Figs.~\ref{fig:alpha_penalty}(b) and (d) show that this value gives essentially the same orbital as the solid green curve for $\alpha_{\rm pen}=7$.

\subsection{Two-electron diatomic molecules}

We now consider a 1D model of a two-electron diatomic molecule. On the two-electron spinfree Hilbert space $\calh \otimes \calh$, the Hamiltonian is
\begin{eqnarray}
\hat{H} = \sum_{i=1}^2 \left( -\frac{1}{2} \frac{\d^2}{\d x_i^2} + v_{\tne}(x_i) \right) + v(x_1 - x_2),
\label{}
\end{eqnarray}
with the same nuclei-electron potential $v_{\tne}(x)$ as in Eq.~\eqref{vneh2+}, and the soft-Coulomb potential in Eq.~\eqref{vSC} is also used for the two-electron interaction $v(x_1 - x_2)$.

We perform configuration-interaction (CI) calculations in the basis of the molecular orbitals $\{\psi_i \}$ of the corresponding one-electron diatomic molecule, obtained from the reference grid calculation, the standard Hermite-Gaussian basis-set calculation, or the discontinuous-Galerkin Hermite-Gaussian basis-set calculations. The ground-state wave function is expanded as
\begin{eqnarray}
\Psi(x_1,x_2) = \sum_{1 \leq i \leq j \leq M} C_{i,j} \Phi_{i,j}(x_1,x_2),
\label{CIexpan}
\end{eqnarray}
where $\Phi_{i,j}$ are spin-singlet symmetrized configurations
\begin{eqnarray}
\Phi_{i,j}(x_1,x_2)=
\begin{cases}\psi_i(x_1)\psi_i(x_2), & i=j,\\[0.3em]
\frac{1}{\sqrt2}\left[\psi_i(x_1)\psi_j(x_2)+\psi_j(x_1)\psi_i(x_2)\right], & i<j.
\end{cases}
\label{Phiij}
\end{eqnarray}
The Hamiltonian matrix in this configuration basis $\{ \Phi_{i,j} \}$ is formed, using the kinetic-energy integrals in Eq.~\eqref{tPint1D} for the case of the discontinuous-Galerkin Hermite-Gaussian basis set, and diagonalized to obtain the ground-state energy $E_\text{CI}$ and the coefficients $C_{i,j}$.

\begin{figure}[t]
\centering
\begin{tabular}{cc}
\includegraphics[width=0.4\linewidth]{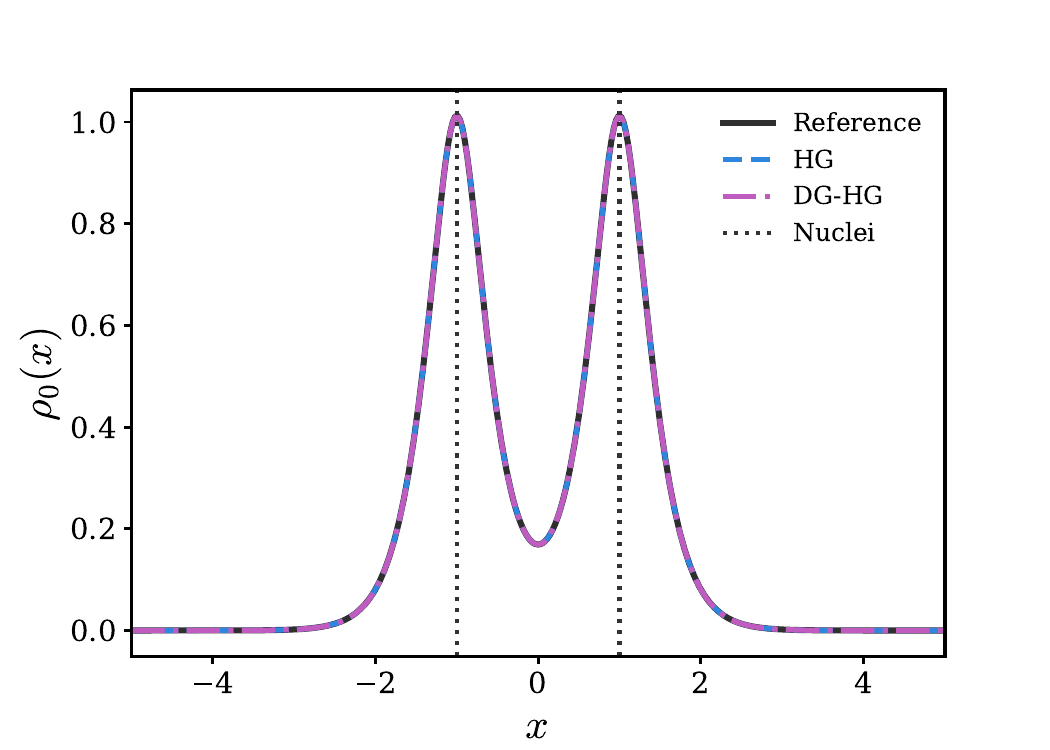} &
\includegraphics[width=0.375\linewidth]{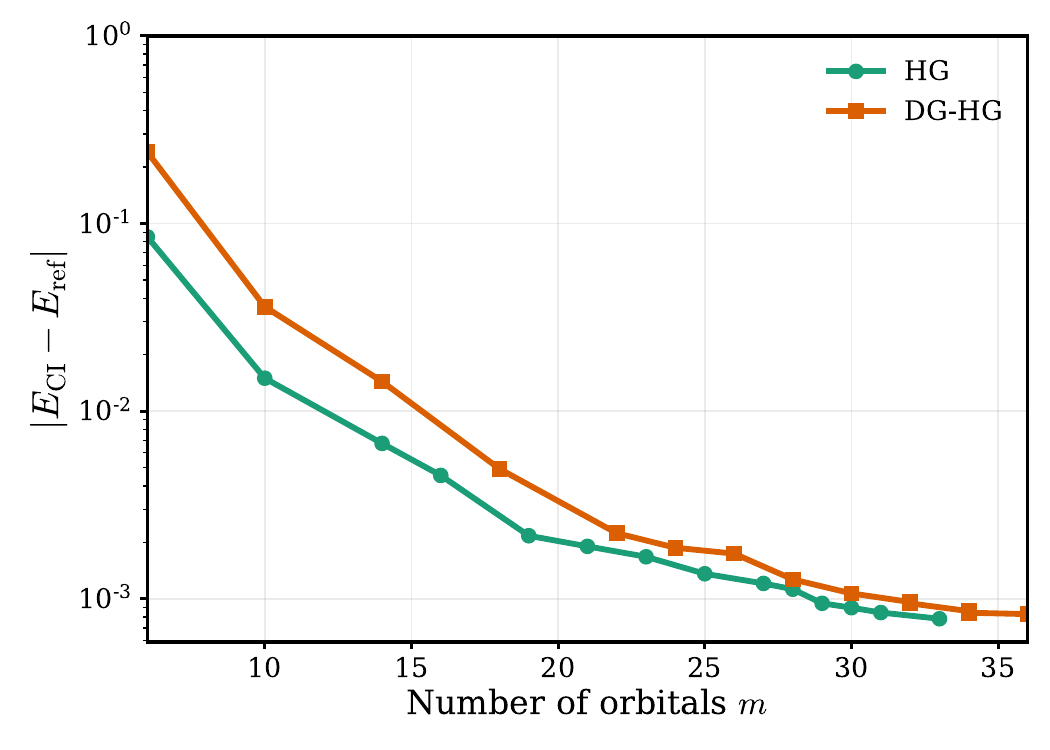}\\[0.4em]
(a) H$_2$ molecule, ground-state density &
(b) H$_2$ molecule, ground-state energy \\[0.8em]
\includegraphics[width=0.4\linewidth]{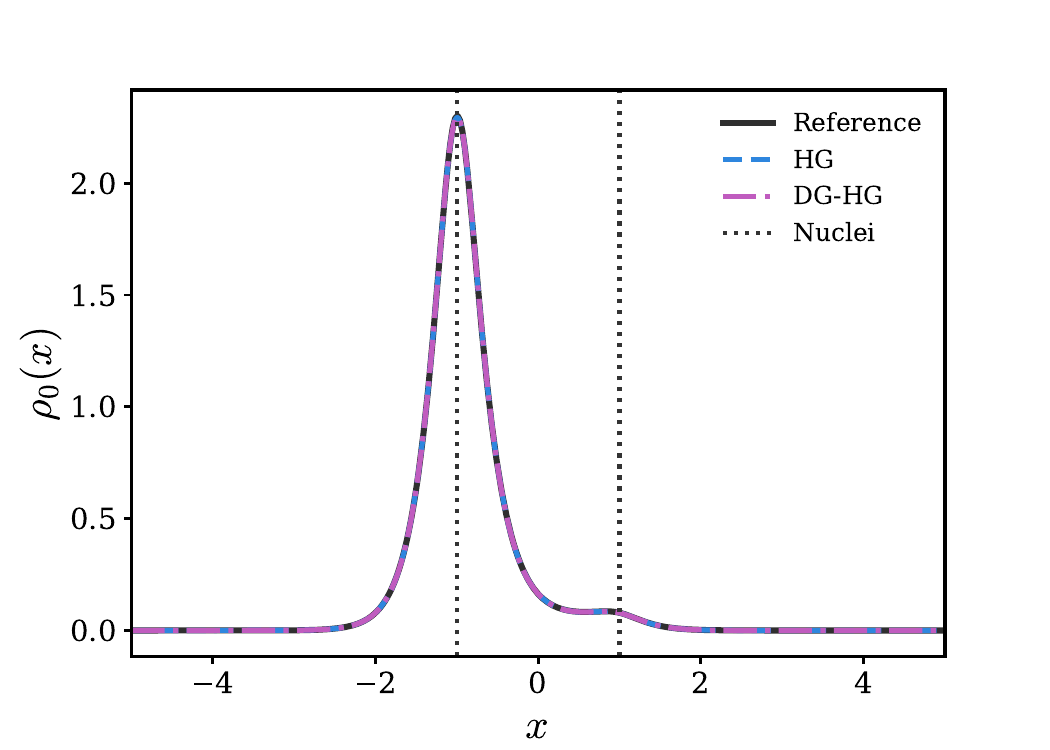} &
\includegraphics[width=0.375\linewidth]{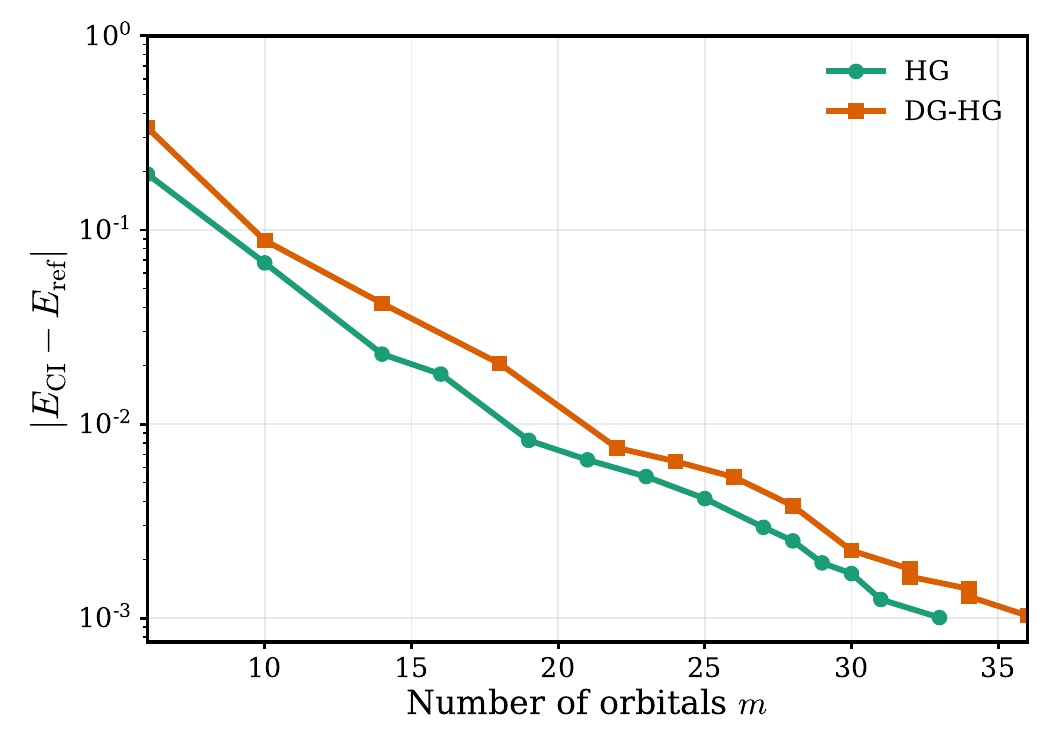}\\[0.4em]
(c) HeH$^+$ molecule, ground-state density &
(d) HeH$^+$ molecule, ground-state energy \\[0.8em]
\end{tabular}
\caption{Ground-state density and convergence of the ground-state energy of the one-dimensional H$_2$ and HeH$^+$ molecules obtained with configuration-interaction (CI) calculations using molecular orbitals from the reference grid method, the standard Hermite-Gaussian (HG) basis set, and the discontinuous-Galerkin Hermite-Gaussian (DG-HG) basis set.}
\label{fig:density_convergence}
\end{figure}

We consider the case of nuclear charges $Z_\L=Z_\R=1$, corresponding to the one-dimensional H$_2$ molecule, and the case of nuclear charges $Z_\L=2$ and $Z_\R=1$, corresponding to the one-dimensional HeH$^{+}$ molecule, both at a fixed internuclear distance of $R=2$.
Figure~\ref{fig:density_convergence} reports the ground-state one-electron density obtained from the CI wave function together with the convergence of the CI energy error as a function of the number of molecular orbitals per atom retained in the CI calculation. All the methods produce nearly identical densities. The discontinuous-Galerkin calculation exhibits a energy convergence with respect to the number of orbitals very comparable to that of the standard basis-set calculation, indicating that the truncation induced by the domain decomposition does not significantly degrade the accuracy of the resulting ground-state energy.

One useful aspect of the strictly localized orbital basis is that the CI expansion in Eq.~\eqref{CIexpan} can be unambiguously decomposed into domain-pair contributions
\begin{eqnarray}
\Psi(x_1,x_2) &=& \sum_{1 \leq \mu \leq \nu \leq m} C_{\L,\mu;\L,\nu} \Phi_{\L,\mu;\L,\nu}(x_1,x_2) + \sum_{1 \leq \mu \leq \nu \leq m} C_{\R,\mu;\R,\nu} \Phi_{\R,\mu;\R,\nu}(x_1,x_2) 
\nonumber\\
&&+ \sum_{1 \leq \mu, \nu \leq m} C_{\L,\mu;\R,\nu} \Phi_{\L,\mu;\R,\nu}(x_1,x_2),
\label{CIexpandecomp}
\end{eqnarray}
where $\Phi_{p,\mu;p',\nu}$ are spin-singlet symmetrized configurations similar to the ones in Eq.~\eqref{Phiij} but expressed in terms of the strictly localized orthonormal orbitals $\{ \tilde{\phi}_{p,\mu} \}_{\mu=1,...,m}^{p=\L,\R}$. Since the configurations $\{\Phi_{p,\mu;p',\nu}\}$ are orthonormal, we can unambiguously define the corresponding domain-pair weights
\begin{eqnarray}
W_{\L\L}=\sum_{1 \leq\mu\leq \nu\leq m}|C_{\L,\mu;\L,\nu}|^2,\qquad
W_{\R\R}=\sum_{1 \leq\mu\leq \nu\leq m}|C_{\R,\mu;\R,\nu}|^2,\qquad
W_{\L\R}=\sum_{1 \leq\mu,\nu\leq m}|C_{\L,\mu;\R,\nu}|^2.
\end{eqnarray}
Assuming the wave function $\Psi$ is normalized, we have $W_{\L\L}+W_{\R\R}+W_{\L\R}=1$. The $\L\R$ sector corresponds to one electron on each domain and can therefore be interpreted as the covalent component of the wave function. The $\L\L$ and $\R\R$ sectors correspond to two electrons localized on the same domain and represent the two ionic components of the wave function. In the symmetric case ($Z_\L=Z_\R$), we have of course $W_{\L\L}=W_{\R\R}$ by symmetry. In the asymmetric case ($Z_\L \not= Z_\R$), the relative magnitude of $W_{\L\L}$ and $W_{\R\R}$ directly informs on the polarization of the bond toward the more attractive nucleus.

\begin{figure}[t]
\centering
\begin{tabular}{cc}
\includegraphics[width=0.4\linewidth]{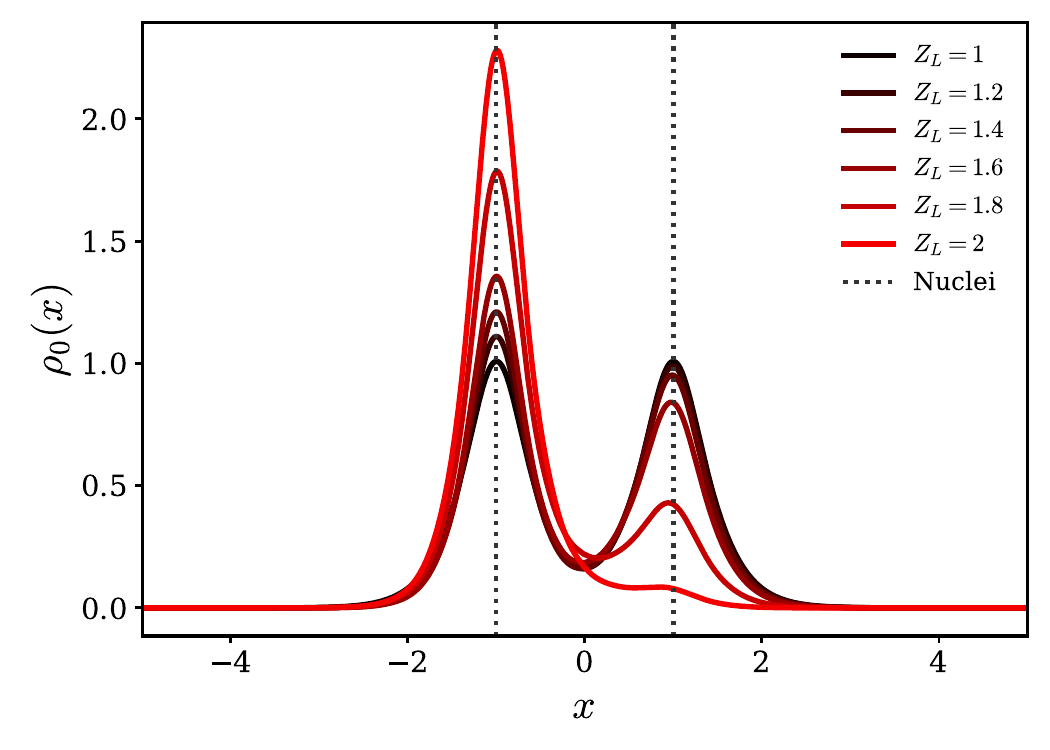} \hspace{1cm} &
\includegraphics[width=0.4\linewidth]{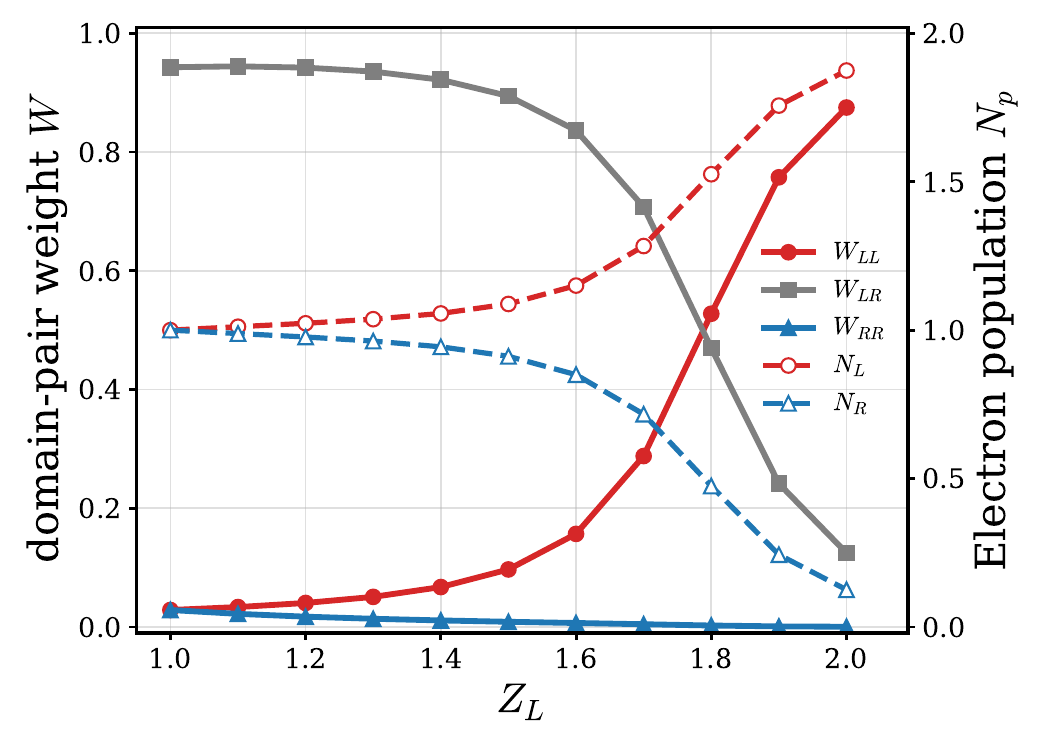}
\end{tabular}
\caption{Evolution of the electronic structure as the left nuclear charge $Z_\L$ is increased from the symmetric H$_2$ case ($Z_\L=Z_\R=1$) toward the asymmetric HeH$^+$ case ($Z_\L=2$ and $Z_\R=1$). The left panel reports the ground-state density, while the right panel reports both the domain-pair weights $W_{\L\L}$ (left ionic), $W_{\R\R}$ (covalent), and $W_{\L\R}$ (right ionic) and the electron populations $N_\L$ and $N_\R$.}
\label{fig:ionicity}
\end{figure}

Figure~\ref{fig:ionicity} shows the evolution of the electronic structure as the left nuclear charge $Z_\L$ is increased from the symmetric H$_2$ case ($Z_\L=Z_\R=1$) toward the asymmetric HeH$^+$ case ($Z_\L=2$ and $Z_\R=1$). The ground-state density is reported, as well as the domain-pair weights $W_{\L\L}$, $W_{\R\R}$, $W_{\L\R}$ and the electron populations $N_\L$ and $N_\R$ on each domain. As expected, as $Z_\L$ increases, the covalent contribution $W_{\L\R}$ decreases, the left-ionic contribution $W_{\L\L}$ increases, and the right-ionic contribution $W_{\R\R}$ becomes negligible. We thus obtain a direct interpretation of the covalent and ionic character of the bond thanks to the discontinuous-Galerkin calculation in the basis of the strictly localized orbitals.

\section{Conclusion}
\label{sec:conclusion}

We have presented a rigorous electronic-structure theory of strictly localized orbitals associated with a spatial partition of the one-electron Hilbert space that remains well defined in the complete basis-set limit. Each strictly localized orbital is supported on a spatial domain and may be discontinuous at domain interfaces. For a QTAIM partition, the effective atomic orbitals introduced by Mayer provide an example of such strictly localized orbitals. Using the interior-penalty discontinuous Galerkin method, these strictly localized orbitals can be employed in variational electronic-structure calculations despite their discontinuities at domain interfaces.

Numerical illustrations on one-dimensional diatomic model systems serve as a proof of concept for the proposed framework. They show that variational calculations can be carried out in a basis of strictly localized orbitals while maintaining good agreement with conventional calculations. The localized representation also enables a direct decomposition of many-electron wave functions into domain-pair contributions, providing a transparent interpretation of the ionic and covalent character of chemical bonds.

In future work, the present approach should be tested on real molecular systems. The present framework of strictly localized orbitals based on spatial domain decomposition could be useful in VB theory~\cite{ShaHib-BOOK-08} and in local-correlation methods~\cite{Nag-CS-24}, and may also offer a natural framework for quantum embedding methods based on spatially localized subsystems~\cite{JonMosSchRat-JACS-20}.

\section*{Acknowledgement}
We thank Eric Cancès et Yvon Maday for insightful discussions.


\end{document}